\documentclass[
onecolumn, 
11pt, 
tightenlines,
superscriptaddress, 
nofootinbib, 
preprintnumbers, 
prd 
]{revtex4-1}
\pdfoutput=1

\usepackage[centering,hmargin=2.5cm,vmargin=2.5cm]{geometry}
\usepackage[colorlinks=true,citecolor=blue,linkcolor=blue]{hyperref}
\usepackage[T1]{fontenc}
\usepackage[normalem]{ulem}
\usepackage{amsmath,amssymb}
\usepackage{mathtools}
\usepackage{epsfig}
\usepackage{graphicx} 
\usepackage{grffile} 
\usepackage{url}
\usepackage{color}
\usepackage{multirow}
\usepackage{placeins}
\usepackage[dvipsnames]{xcolor}
\usepackage{epstopdf}
\usepackage{siunitx}
\usepackage{enumitem}
\usepackage{cleveref}
\usepackage{lipsum}
\usepackage{gensymb}
\usepackage{xspace}
\usepackage{titlesec} 
\usepackage{booktabs} 
\usepackage{subfigure}
\usepackage[subfigure,titles]{tocloft} 

\allowdisplaybreaks

\makeatletter

\renewcommand{\p@subsection}{}

\renewcommand{\p@subsubsection}{}
\makeatother

\titleformat*{\section}{\centering\bfseries\scshape}
\titlelabel{\thetitle\quad}
\titleformat*{\paragraph}{\bfseries}
\titlespacing*{\paragraph}{0pt}{3.25ex plus 1ex minus .2ex}{1em}

\newcommand{\bra}[1]{\ensuremath{\langle #1 |}}   
\newcommand{\ket}[1]{\ensuremath{| #1 \rangle}}   
\newcommand{\ev}[1]{\ensuremath{\left\langle #1 %
        \right\rangle}} 

\renewcommand{\vec}[1]{{\boldsymbol{#1}}}

\newcommand{\pd}[2]{\frac{\partial #1}{\partial #2}}                  
\newcommand{\pdtext}[2]{{\partial #1} / {\partial #2}}                
\newcommand{\td}[2]{\frac{\mathrm d #1}{\mathrm d #2}}                
\newcommand{\tdtext}[2]{{\mathrm d #1}/ {\mathrm d #2}}               
\newcommand{\upd}{\mathrm d}                                          
\newcommand{\ba}[1]{\bigg(#1\bigg)}                                   
\newcommand{\bb}[1]{\bigg[#1\bigg]}                                   

\newcommand{\f}{\ensuremath{f_\chi}\xspace}
\newcommand{\A}{\ensuremath{\mathcal A}\xspace}
\newcommand{\feq}{\ensuremath{\f^\text{eq}}\xspace}


\begin{document}

\title{Detailed Calculation of Primordial Black Hole Formation \\
       During First-Order Cosmological Phase Transitions}

\author{Michael J.\ Baker}
\email{michael.baker@unimelb.edu.au}
\affiliation{ARC Centre of Excellence for Dark Matter Particle Physics, 
School of Physics, The University of Melbourne, Victoria 3010, Australia}

\author{Moritz Breitbach}
\email{breitbach@uni-mainz.de}
\affiliation{PRISMA Cluster of Excellence \& Mainz Institute for
    Theoretical Physics, \\
    Johannes Gutenberg University, Staudingerweg 7, 55099
    Mainz, Germany}

\author{Joachim Kopp}
\email{jkopp@cern.ch}
\affiliation{PRISMA Cluster of Excellence \& Mainz Institute for
    Theoretical Physics, \\
    Johannes Gutenberg University, Staudingerweg 7, 55099
    Mainz, Germany}
\affiliation{Theoretical Physics Department, CERN,
    Esplanade des Particules, 1211 Geneva 23, Switzerland}

\author{Lukas Mittnacht}
\email{lmittna@uni-mainz.de}
\affiliation{PRISMA Cluster of Excellence \& Mainz Institute for
    Theoretical Physics, \\
    Johannes Gutenberg University, Staudingerweg 7, 55099
    Mainz, Germany}

\date{\today}

\preprint{CERN-TH-2021-113}
\preprint{MITP-21-036}


\begin{abstract}
    \noindent
    Primordial black holes could potentially form during a first-order cosmological phase transition due to a build-up of particles which are predominantly reflected from the advancing bubble walls. After discussing the general mechanism, we examine the criteria that need to be satisfied for a black hole to form.  We then set out the Boltzmann equation that describes the evolution of the relevant phase space distribution function, carefully describing our treatment of the Liouville operator and the collision term.  Assuming a spherical false vacuum pocket of sufficient size and a constant wall velocity, we find that black holes can form in a range of different scenarios.
\end{abstract}

\maketitle

\newcommand{\contentsname}{}
\setlength{\cftbeforesecskip}{3pt}
\setlength{\cftsecnumwidth}{1.5em}
\setlength{\cftsubsecindent}{\cftsecnumwidth}
\setlength{\cftsubsecnumwidth}{2.2em}
\setlength{\cftsubsubsecindent}{\cftsecnumwidth+\cftsubsecnumwidth}
\setlength{\cftsubsubsecnumwidth}{2.9em}
\renewcommand{\cftdotsep}{2}
\tableofcontents

\section{Introduction}
\label{sec:introduction}

The possibility that our Universe contains a population of primordial black holes (PBHs) -- black holes formed during the Universe's infancy just after the Big Bang -- has recently received tremendous attention. While this recent surge in interest has mostly been sparked by the advent of gravitational-wave astronomy and thus of a new tool to search for black holes \cite{Abbott:2016blz}, there is a broad theoretical motivation for considering the potential existence of primordial black holes.  In particular, depending on their mass and abundance, primordial black holes could constitute the dark matter (DM) in the Universe~\cite{%
Carr:2016drx,                  
Green:2020jor,                 
Carr:2020xqk,                  
Villanueva-Domingo:2021spv,    
Katz:2018zrn},                  
produce it through Hawking radiation~\cite{
Green:1999yh,         
Khlopov:2004tn,       
Fujita:2014hha,
Allahverdi:2017sks,
Lennon:2017tqq,       
Morrison:2018xla,     
Hooper:2019gtx,       
Masina:2020xhk,       
Baldes:2020nuv,       
Gondolo:2020uqv,      
Bernal:2020kse,       
Bernal:2020bjf}        
or asymmetrically reheat the dark and visible sectors~\cite{
Sandick:2021gew}.      
They can modify the expansion history of the Universe~\cite{%
Hooper:2019gtx,       
Chaudhuri:2020wjo},   
destroy unwanted monopoles and domain walls~\cite{
Stojkovic:2004hz,
Stojkovic:2005zh},
evolve into the supermassive black holes found at centre of most galaxies~\cite{
Bean:2002kx},
or seed large scale structure formation~\cite{Hoyle:1966,Ryan:1972,Carr:1984,Afshordi:2003zb,Carr:2020gox}.

Similarly, the physics of early Universe phase transitions has become a popular topic in recent years, again driven by the prospect of observing gravitational waves from these phase transitions in upcoming experiments~\cite{
  Hogan:1983ixn,
  Witten:1984rs,
  Turner:1990rc,
  Schwaller:2015tja,  
  Caprini:2015zlo,    
  Caprini:2019egz,    
  Hindmarsh:2020hop}. 
Besides significant advances in our understanding of the dynamics of these phase transitions \cite{
  Bodeker:2009qy,     
  Bodeker:2017cim,    
  Espinosa:2019hbm,   
  Mukhanov:2021ggy,   
  Espinosa:2021qeo}   
and in our computational techniques \cite{
  Wainwright:2011kj,  
  Masoumi:2017trx,    
  Espinosa:2018hue,   
  Espinosa:2018voj,   
  Espinosa:2018szu,   
  Guada:2018jek,      
  Guada:2020xnz},     
phase transitions have been widely used as phenomenological tools, for instance to explain the baryon asymmetry of the Universe via electroweak baryogenesis \cite{
    Kuzmin:1985mm,        
    Shaposhnikov:1986jp,  %
    Shaposhnikov:1987tw,  %
    Carena:1996wj,
    Riotto:1998zb,        
    Huber:2006wf,         
    Cline:2006ts,         
    Morrissey:2012db,     
    Vaskonen:2016yiu,     
    Garbrecht:2018mrp,    
    deVries:2018tgs,      
    Cline:2020jre,        
    Fuchs:2020pun}        
and other mechanisms \cite{
    Dutta:2006pt,         
    Shelton:2010ta,       
    Dutta:2010va,         
    Hall:2019ank,         
    Cline:2017qpe,        
    Huang:2017kzu,
    Arakawa:2021wgz},     
to form DM ``nuggets'' \cite{
  Hong:2020est,       
  Gross:2021qgx,      
  Asadi:2021pwo},
or to set the abundance of particle dark matter \cite{
  Baker:2016xzo,      
  Baker:2017zwx,      
  Baker:2018vos,      
  Baker:2019ndr}.     
Most recently, several groups, including the authors of this paper, have explored the possibility that matter compressed by the advancing bubble walls in a first-order phase transitions may collapse into black holes \cite{Gross:2021qgx, Baker:2021nyl, Kawana:2021tde}. These mechanisms offer an interesting alternative to PBH production via the more widely studied collapse of density perturbations created during inflation \cite{
  Carr:1975qj,
  Ivanov:1994pa,
  Garcia-Bellido:1996mdl,
  Silk:1986vc,
  Kawasaki:1997ju,
  Yokoyama:1995ex,
  Pi:2017gih},
a mechanism that requires special, in particular very flat, inflaton potentials. Other proposed PBH production mechanisms include the collapse of topological defects~\cite{
  Hawking:1987bn,
  Polnarev:1988dh,
  MacGibbon:1997pu,
  Rubin:2000dq,
  Rubin:2001yw,
  Ashoorioon:2020hln,
  Brandenberger:2021zvn}
or scalar condensates~\cite{
  Cotner:2016cvr,
  Cotner:2019ykd},
as well as collisions of bubble walls during a first-order phase transition~\cite{
  Crawford:1982yz,
  Kodama:1982sf,
  Moss:1994pi,          
  Freivogel:2007fx,     
  Hawking:1982ga,       
  Johnson:2011wt,       
  Kusenko:2020pcg}
  or via delayed vacuum decay~\cite{
  Liu:2021svg}.

In the present paper, we expand on the mechanism proposed in ref.~\cite{Baker:2021nyl} which posits that at some temperature $T_n$ the early Universe underwent a strong first-order phase transition, and that during this phase transition the mass of a particle species $\chi$ changed substantially from being nearly massless ($m_\chi \approx 0$) in the false vacuum phase to being very massive ($m_\chi \gg T_n$) in the true vacuum phase.  Due to this large gain in mass, very few $\chi$ particles were able to enter the true vacuum phase (due to energy conservation).  Instead, they were pushed ahead of the bubble walls that separate the two phases, such that towards the end of the phase transition their energy density within the remaining false vacuum pockets was significantly enhanced compared to their thermal equilibrium energy density. In some of the pockets, it may have become large enough to trigger black hole formation.

In the following we describe in detail the numerical calculations we carried out to simulate this PBH formation mechanism. After setting down our notation, introducing a simple toy model, and discussing the mechanism qualitatively in \cref{sec:general mechanism,sec:pt-properties}, we describe the various criteria for the formation of black holes in \cref{sec:Schwarzschild-criteria,sec:JI-criteria}.  Either the pocket, or bubble, containing the $\chi$ overdensity must shrink below its own Schwarzschild radius, or it has to satisfy the Jeans instability criteria to initiate collapse, while at the same time the collapse must not be stopped by either degeneracy pressure or by the outward pressure that $\chi$ exerts on the wall.  In \cref{sec:boltzmann} we derive the Boltzmann equation that describes the dynamics of $\chi$ during the phase transition, and we explain how we solve it numerically.  We present our results in \cref{sec:results} and discuss the dependence on the phase transition properties in \cref{sec:discussion}. We conclude in \cref{sec:conclusions}.

\section{Black Hole Formation During Cosmological First-Order Phase Transitions}
\label{sec:mechanism}

\subsection{General Mechanism}
\label{sec:general mechanism}

To illustrate the mechanism of PBH formation during a first-order phase transition, we introduce a minimal toy model which extends the Standard Model (SM) with a Dirac fermion $\chi$ and a real scalar field $\phi$. Both $\chi$ and $\phi$ are gauge singlets. Their Lagrangian contains a Yukawa interaction and the scalar potential,
\begin{align}
  \mathcal L \supset - y_\chi \phi\overline{\chi}{\chi} - V(\phi,H) \,,
  \label{eq:L}
\end{align}
where $y_\chi$ is the coupling constant (assumed to be real for simplicity), and $V(\phi,H)$ is the scalar potential which depends on $\phi$ and the SM Higgs field $H$. To remain as general as possible, we do not assume a specific potential and only assume a few properties which are necessary for our purpose:
\begin{enumerate}
    \item The phase transition, occurring at nucleation temperature $T_n$, must be first order to ensure that the true and false vacuum phases coexist at the same time and that they are separated by domain walls.
    
    \item The vacuum expectation value (vev) of $\phi$, $\ev{\phi} \equiv \bra{0}\phi\ket{0}$ is $\ll T_n$ above the temperature $T_n$ (we also assume that $m_\chi \ll T_n$ in this regime), while $\ev{\phi} \gg T_n$ after the transition (giving $\chi$ a mass $m_\chi \approx y_\chi\!\ev{\phi} \gg T_n$). This ensures that the domain walls separating the two phases are all but impermeable to $\chi$ particles.
    
    \item The phase transition should proceed very slowly.  That is, the time derivative of the bounce action should be small. This feature is often realised in supercooled phase transitions \cite{DelleRose:2019pgi} (see also refs.~\cite{Konstandin:2011dr, Jaeckel:2016jlh, Hambye:2018qjv}).
\end{enumerate}

\begin{figure}
    \centering
    \includegraphics{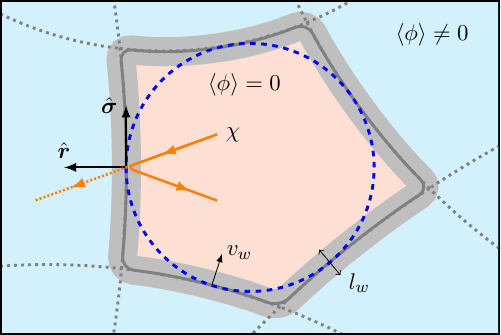}
    \caption{
      A schematic picture of the first-order cosmological phase transition: regions of true vacuum (blue) are expanding with speed $v_w$ and have coalesced, resulting in a shrinking, approximately-spherical bubble of false vacuum (light red).  Very high momentum $\chi$ particles pass through the bubble wall and convert kinetic energy into a large mass, while lower momentum $\chi$ particles are reflected due to energy conservation.  The reflected $\chi$ particles build up inside the bubble, creating an energy overdensity which may lead to a black hole.  The local coordinate system and the bubble wall thickness, $l_w$, are also shown.}
    \label{fig:illustration}
\end{figure}

First-order phase transitions are triggered by the nucleation of true vacuum bubbles in a Universe that is still in the false vacuum. The bubbles expand since their walls are driven forward by the latent heat release associated with the phase transition (that is, the effective potential difference between the false and true vacua of $\phi$). The bubbles collide and merge until the whole Universe has transitioned to the true vacuum state.  As we will see, in our scenario it is more appropriate to describe the system of interest as a shrinking false-vacuum bubble rather than a population of expanding true-vacuum bubbles as in most other papers on phase transitions. Such a shrinking bubble is illustrated by the light-red region in \cref{fig:illustration}.

We will assume that at the onset of the phase transition, there is an equilibrium population of relativistic $\chi$ particles that is not in thermal contact with the SM bath.\footnote{Here and in the following, $\chi$ is meant to refer to both particles and antiparticles, except when we refer to $\bar\chi$ explicitly.}  These particles could have been in thermal contact at a higher temperature and, as they are still relativistic, would have maintained the same temperature as the SM bath as the Universe expanded and they decoupled.  Since $\chi$ particles have a typical kinetic energy $\sim T_n$ at the time of the phase transition, and since their mass would change by an amount $\gg T_n$ if they traversed the wall, conservation of energy implies that most $\chi$ particles are reflected from the wall (solid orange path in \cref{fig:illustration}).  This means that an overabundance of $\chi$ will form inside the shrinking bubble. A $\chi$ particle also receives a kick on reflection from the wall, increasing its momentum.  We will show that the energy overdensity inside the shrinking bubble, due to both the $\chi$ overabundance and their increased momenta, may become sufficiently large to form a black hole.

Note that in our mechanism a shrinking bubble can only yield a black hole if no new bubbles nucleate in its interior, up to the point when the black hole forms. Otherwise, the overdense regions would be split up, and each of these smaller regions may not obtain a large enough overdensity to trigger black hole formation.  For any given region, the probability of producing a black hole is typically tiny because of this constraint.  However, even if black hole formation is rare, it can still lead to a sizeable PBH abundance today~\cite{Baker:2021nyl}.

\subsection{Properties of the Phase Transition}
\label{sec:pt-properties}

Since we do not specify the scalar potential, we will describe the phase transition by its overall properties: the bubble nucleation temperature $T_n$, the scalar vev in the true vacuum $\ev{\phi}^\infty$,\footnote{Since we will consider $\ev{\phi}$ as a function of position near the wall, we use $\ev{\phi}^\infty$ to denote its value deep inside the true vacuum.} the wall velocity $v_w$ and the wall thickness $l_w$.  In our numerical calculation we make several simplifying assumptions:
\begin{enumerate}
    \item We assume that the portal couplings $\phi |H|^2$ and $\phi^2 |H|^2$ are small enough so the phase transition in $\phi$ can be treated independently of the electroweak phase transition, but large enough that $\phi$ remains in thermal and chemical equilibrium throughout the phase transition.  When the phase transition occurs at temperatures greater than $\sim \SI{1}{GeV}$, portal couplings as small as $10^{-3}$ are sufficient to keep $\phi$ in thermal contact with the SM bath~\cite{Baker:2017zwx}.  If the phase transition occurs at temperatures $\gg \SI{1}{TeV}$, these portal couplings may give large contributions to the SM Higgs mass parameter.  However, smaller portal couplings are also viable if $\phi$ has couplings to additional new particles that keep it in equilibrium.
    
    \item The shrinking region is spherically symmetric.  While this may not be the case to begin with, surface tension in the bubble walls will tend to pull the bubble into a sphere so it will become a better approximation as time passes.
    
    \item The latent heat of the phase transition is large enough to sufficiently compress the $\chi$ particles, but less than the energy density of the thermal bath.  This ensures that the Universe remains radiation-dominated throughout the phase transition, and we do not need to take into account a possible intermittent phase of inflation.
    
    \item The wall profile, velocity $v_w$ and thickness $l_w$, as well as the temperature $T_n$, are constant throughout the phase transition.
\end{enumerate}
These assumptions do not need to be satisfied for the general mechanism to work.  However, relaxing them will typically require a more detailed analysis than we perform here.  We will discuss assumptions 3.~and 4.~in detail in \cref{sec:discussion} (in particular showing that it is conservative to assume that the velocity is constant) .

\subsection{Forming a Primordial Black Hole -- The Schwarzschild Criteria}
\label{sec:Schwarzschild-criteria}

The most straightforward condition for black hole formation is that an energy overdensity is contained in a region smaller than its Schwarzschild radius. In this case, the gravitational force which the overdensity exerts on a test particle is so great that an event horizon forms. A test particle entering this horizon will never be able to escape.  Note that it is the energy \emph{overdensity} that is relevant here, not the total energy contained in the region, because a homogeneous background density does not cause a net gravitational force on a test particle.  In this section we will determine the $\chi$ energy density required to satisfy this condition.  Since the time scale of gravitational collapse can be significantly shorter than the phase transition completion time, in \cref{sec:JI-criteria} we will also consider several conditions that together would lead to an overdensity of particles collapsing into a black hole in the absence of the bubble wall: the internal pressure must be weak enough so as not to prevent gravitational collapse (the Jeans instability criteria), Fermi degeneracy pressure must not prevent collapse (as for white dwarf stars below the Chandrasekhar limit), and the particles must have a way of shedding angular momentum efficiently.  We will see that the Fermi degeneracy pressure criteria is typically satisfied only after the region has become smaller than its Schwarzschild radius, while the rate of angular momentum shedding depends on the $\chi$ interaction strength.  We will therefore use the Schwarzschild criteria outlined in this section as our main criteria for determining black hole formation.

The spherical region containing the energy overdensity is smaller than its Schwarzschild radius  when
\begin{align}
    r_w(t) \qquad < \qquad
    r_s \equiv 2 G E_\text{tot}^{(<r_w)} \,,
    \label{eq:rs}
\end{align}
where $E_\text{tot}^{(<r_w)}$ is the energy overdensity contained in a bubble of radius $r_w(t)$ at time $t$.  Conservatively taking only the energy density carried by $\chi$ into account (and not that, for instance, in the bubble wall itself), a black hole thus forms when\footnote{Since $\rho_\chi \approx 0$ outside the bubble, we approximate $\Delta\rho_\chi(t)$ as the $\chi$ energy density inside the bubble, $\rho_\chi(t)$.}
\begin{align}
    \Delta\rho_\chi(t) \approx \rho_\chi(t) > \frac{\pi^2}{30} g_\star\,T_n^4 \bigg(\frac{r_H^0}{r_w(t)}\bigg)^{\!2} \,,
    \label{eq:schwarzschild}
\end{align}
where $g_\star$ is the total effective number of relativistic degrees of freedom in the Universe at temperature $T_n$, and where we have introduced the Hubble radius at an initial time $t_0$ via the Friedmann equation,
\begin{align}
    r_H^0 \equiv \frac{1}{H(t_0)} = \sqrt{\frac{3}{8\pi G} \frac{30}{\pi^2 g_\star}}
                        \frac{1}{T_n^2} \,.
\end{align}
Note that since the phase transition does not complete instantaneously, the Hubble radius will be somewhat larger at the time of black hole formation than at $t_0$.  To track physical distances, we write them in terms of the Hubble radius at this time, $r_H^0$.

Although the Friedmann equation implies that the temperature will also reduce during the phase transition, we neglect this effect since $T_n \propto \sqrt{H}$ in a radiation-dominated universe and we find that if black holes form, they do so after not much more than one Hubble time.

We can get an idea of how much a bubble needs to shrink to satisfy the Schwarzschild criteria by using a naive relation between energy density and bubble radius.  Assuming that no $\chi$ particles pass through the bubble wall, the number density would be proportional to $[r_w^0/r_w(t)]^3$, where $r_w^0\equiv r_w(t_0)$ is the initial bubble radius.  The $\chi$ particles gain kinetic energy each time they are reflected from the bubble wall, where $\upd E = 2 v_w E$ for non-relativistic wall velocities. The time between reflections for a purely radial $\chi$ trajectory is $\upd t = 2 r_w(t) = 2 (r_w^0 - v_w t)$. Solving $\int \upd E/E = \int \upd t \, v_w / r_w(t) = -\!\int \upd r_w / r_w$, we find that $E \propto r_w^0/r_w(t)$.  Taking these two effects together, we would expect the energy density to roughly scale as $[r_w^0/r_w(t)]^4$.  Taking the $\chi$ particles to initially be in equilibrium, $\rho_\chi(t_0) = \rho_\chi^\text{eq} = (7 \pi^2/240) g_\chi T_n^4$, we then obtain the $\chi$ energy density as a function of time,
\begin{align}
    \label{eq:rho-chi}
    \rho_\chi(t) \approx \bigg( \frac{r_w^0}{r_w(t)} \bigg)^{\!4}
                     \frac{7\pi^2}{240} g_\chi T_n^4 \,,
\end{align}
where $g_\chi=4$ is the number of degrees of freedom of $\chi+\bar\chi$.  We can use this result to express the Schwarzschild criteria as a condition on the bubble radius, 
\begin{align}
\label{eq:schwarzschild-crit}
    \frac{r_w(t)}{r_H^0} &\lesssim\, \sqrt{\frac{7}{8} \frac{g_\chi}{g_\star}} \bigg(\frac{r_w^0}{r_H^0}\bigg)^{\!2} \,.
\end{align}
Interestingly, this criteria is independent of any physical scale, except through $g_\star$.  As an example, if $r_w^0 = 1.5\,r_H^0$ and $T_n \gtrsim \SI{100}{GeV}$, the Schwarzschild condition is satisfied once $r_w(t)\approx 0.4\,r_H^0$.  That is, after the bubble radius has decreased by a factor of just 3.7.  Note that for bubbles this large, the Schwarzschild criteria is satisfied soon after the bubble has entered the Hubble horizon.  In this case, black hole formation will not be immediate due to causality, which implies that a test particle cannot be affected by the whole energy overdensity until a signal travelling at the speed of light has had enough time to traverse the bubble. Still, once the Schwarzschild criteria is satisfied, black hole formation is practically unavoidable, so we can safely use \cref{eq:schwarzschild} as our main black hole formation criteria.

\subsection{Forming a Primordial Black Hole -- The Jeans Instability Criteria}
\label{sec:JI-criteria}

\subsubsection{Jeans Instability}
\label{sec:jeans-instability}

In principle, black holes can also form via gravitational collapse, that is via a Jeans instability. However, we will ultimately find that this mechanism is not relevant in the scenarios we consider.  We here discuss the calculations that support this conclusion. The Jeans instability criteria requires that the free-fall time for a spherically symmetric volume falls below the sound-crossing time,
\begin{align}
  t_\text{ff} \equiv \sqrt{\frac{3\pi}{32 G \Delta\rho_\chi(t)}}
  \qquad < \qquad
  t_\text{s} \equiv \frac{r_w(t)}{c_\text{s}} \,,
\end{align}
where $G$ is the gravitational constant, $c_\text{s}=\sqrt{1/3}$ is the speed of sound for ultra-relativistic particles, and $\Delta\rho_\chi(t)$ is the local energy overdensity with respect to the surrounding plasma.  Using the Friedmann equation to introduce the Hubble radius, the Jeans instability criteria then becomes
\begin{align}
    \Delta\rho_\chi(t) \approx \rho_\chi(t) > \frac{\pi^4}{360}g_\star\,T_n^4 \bigg(\frac{r_H^0}{r_w(t)}\bigg)^{\!2} \,. 
    \label{eq:Jeans-rho}
\end{align}

Comparing with \cref{eq:schwarzschild}, we see that a black hole may be formed via collapse when the energy density is just a factor $12/\pi^2 \approx 1.2$ smaller than the energy density where the Schwarzschild criteria is satisfied.  

Rephrased as a condition on the bubble radius, \cref{eq:Jeans-rho} can be rewritten as
\begin{align}
    \frac{r_w(t)}{r_H^0} &\lesssim\, 
            \frac{\sqrt{12}}{\pi} 
            \sqrt{\frac{7}{8} \frac{g_\chi}{g_\star}}
            \bigg(\frac{r_w^0}{r_H^0}\bigg)^{\!2} \,,
\end{align}
where we have assumed that $\rho_\chi$ scales with $r_w^4$, based on the arguments given above \cref{eq:rho-chi}.

\subsubsection{Degeneracy Pressure}
\label{sec:degeneracy}

As $\chi$ is a fermion, gravitational collapse into a black hole may be inhibited by degeneracy pressure.\footnote{While gravitational collapse may also be inhibited by slow angular momentum shedding, we do not investigate this further since we find that degeneracy pressure alone inhibits collapse at the parameter points we consider in \cref{sec:results}.} We formalise this condition by computing the total energy of the system,
\begin{align}
    E_\text{tot} = E_\text{grav} + E_\text{kin} \,,
\end{align}
where $E_\text{grav}$ is the total gravitational energy and $E_\text{kin}$ is the kinetic energy of the relativistic $\chi$ gas.  For ultra-relativistic particles occupying a sphere of radius $R$,
\begin{align}
    E_\text{grav} = -\frac{3}{5} \frac{G E_\text{kin}^2}{R} \,.
\end{align}
Note that $R$ does not need to be equal to the bubble radius $r_w(t)$: collapse could in principle start at $R = r_w(t)$ and then be stopped when the population of $\chi$ particles reaches a smaller radius $R < r_w(t)$.  We emphasise that $E_\text{kin}$ and $E_\text{grav}$ are both functions of $R$.  The system will be stable if $\upd E_\text{tot} / \upd R$ is negative because then a decrease in the radius $R$ leads to an increase in energy. If, on the other hand,
\begin{align}
    \td{E_\text{tot}}{R} > 0 \,,
\end{align}
further collapse is energetically favourable.

We can make further analytical progress once the dark matter forms a degenerate Fermi gas: in this case, $E_\text{kin}$ can be calculated by multiplying the phase space distribution function (which is just a step function in the degeneracy limit) by the energy (which equals the momentum for $m_\chi \approx 0$), integrating over momentum, and multiplying by the volume of the bubble. The result is
\begin{align}
    E_\text{kin} = \frac{3}{4} \left( \frac{9 \pi}{2 g_\chi} \right)^{\!1/3} \frac{N_\chi^{4/3}}{R} \,,
\end{align}
where $N_\chi$ is the total number of particles collapsing. The condition $\upd E_\text{tot} / \upd R > 0$ then becomes
\begin{align}
    \td{E_\text{tot}}{R} = \frac{9}{5} \frac{G E_\text{kin}^2}{R^2} - \frac{E_\text{kin}}{R} &> 0 \,,
\intertext{or, equivalently,}
    \frac{G N_\chi^{4/3}}{R^2 g_\chi^{1/3}} > 0.31 \,.
    \label{eq:chandrasekhar-1}
\end{align}
We can now use $N_\chi = \frac{3}{4} [\zeta(3) / \pi^2] g_\chi T_n^3 \cdot \frac{4}{3} \pi (r_w^0)^3$ (assuming that loss of $\chi$ particles through the bubble wall is negligible), the Friedmann equation $r_H^0 \approx 1 / [1.66 \sqrt{g_\star G} T_n^2]$ and $g_\star \sim 100$ to write the collapse condition as
\begin{align}
    g_\chi \bigg( \frac{r_w^0}{r_H^0} \bigg)^{\!4}\bigg( \frac{r_H^0}{R} \bigg)^{\!2} \gtrsim \num{3e2} \,.
    \label{eq:degeneracy-1}
\end{align}
This condition is most stringent at the largest $R$, when $\chi$ particles first form a degenerate Fermi gas. Typically, this point is reached some time after collapse has started, but to be conservative, we will set $R = r_w(t)$ when evaluating \cref{eq:degeneracy-1} in \cref{sec:results}.

We see that the degeneracy criteria is difficult to satisfy.  A bubble with an initial size $r_w^0 \approx r_H^0$ needs to shrink by a factor of around 10. By the time this happens, it will either have accumulated a very large $\chi$ overdensity, or lots of $\chi$ particles will have been lost through the bubble wall or to annihilation. In the former case, the enormous pressure generated by the $\chi$ overdensity would need to be compensated by an equally large latent heat release from the phase transition (see also \cref{sec:latent-heat} below). In the latter case, the loss of $\chi$ particles will reduce the energy overdensity and preclude black hole formation. The difficulty in satisfying \cref{eq:degeneracy-1} while simultaneously maintaining an overdensity large enough to trigger a Jeans instability is the main reason why we find, in \cref{sec:results}, that black holes form due to the bubble shrinking below its Schwarzschild radius (\cref{eq:schwarzschild}) rather than via a Jeans instability.

\section{The Boltzmann Equation}
\label{sec:boltzmann}

To determine whether it is possible to form black holes via our proposed mechanism, we need to calculate the energy density of the $\chi$ particles as a function of space and time during the phase transition.  The energy density is given by the integral over momentum of the phase space distribution function of $\chi$, $\f(\vec{x}, \vec{p}, t)$, multiplied by the energy and the number of degrees of freedom of $\chi$ ($g_\chi=4$ for a Dirac fermion). 

As discussed above, we consider the scenario where the remaining region of false vacuum (where $\ev{\phi} = 0$) can be approximated as spherical, see \cref{fig:illustration}. As a consequence, \f possesses a spherical symmetry and depends on only one spatial coordinate: the distance $r$ to the centre of the spherical region. The momentum dependence of $\f$, on the other hand, cannot be reduced to one dimension: we have to keep track of the radial component $p_r$ and the tangential component $p_\sigma$ separately. Due to the spherical symmetry, a given $\chi$ particle will stay within a plane (ignoring collisions), so the third momentum variable, which indicates the orientation of that plane, can be dropped.  Overall, we therefore write $\f \equiv \f(r, p_r, p_\sigma, t)$.\footnote{Note that while we take a spherical coordinate system in real space, the coordinate system in momentum space is cylindrical.  Our choice of basis in momentum space depends on the position in real space, so that the radial direction in coordinate space aligns with the axial direction in the cylindrical system in momentum space. The momentum component in that direction is $p_r$, while $p_\sigma$ is the momentum component orthogonal to it.}

To calculate the evolution of the phase space distribution function, we start from the general form of the Boltzmann equation,
\begin{align}
    \label{eq:boltzmann}
    \boldsymbol{\mathrm L}[\f] &= \boldsymbol{\mathrm C}[\f] \,,
\end{align}
where the Liouville operator $\boldsymbol{\mathrm L}[\f]$ captures the kinetic evolution of $\chi$, while the collision term $\boldsymbol{\mathrm C}[\f]$ describes scattering and annihilation.
Both terms will be discussed in detail below.

For our calculations, it will be useful to express the phase space distribution function in terms of the equilibrium Fermi--Dirac distribution, i.e.,
\begin{align}
    \label{eq:f-ansatz}
    \f(r, p_r, p_\sigma, t) &\equiv \A(r, p_r, p_\sigma, t) \,\feq(r, p_r, p_\sigma, t) \,,
    \intertext{with}
    \feq(r, p_r, p_\sigma, t) &\equiv \bigg[1 + {\exp}\bigg(\frac{(m^2_\chi(r,t) + p_r^2 + p_\sigma^2)^{1/2}}{T_n} \bigg)\bigg]^{-1}\,,
\end{align}
and where $T_n$ is the temperature of the thermal bath at the time of the phase transition. Note that we define $\f$ and $\feq$ to be the phase space distributions of a single degree of freedom of $\chi$. In \cref{eq:f-ansatz}, we introduce the \textit{phase space enhancement factor} \A, where $\A=1$ corresponds to an equilibrium distribution.  \A measures the departure of $\chi$ from equilibrium induced by interactions with the bubble wall, and it will be the main subject of our numerical calculations.

The change in $m_\chi$ across the bubble wall is modelled as
\begin{align}
    m_\chi(r,t) &= y_\chi \ev{\phi}\!(r,t) \,,
    \label{eq:mass-profile}
\end{align}
where the wall profile -- that is, the coordinate- and time-dependent vev of $\phi$ -- is given by
\begin{align}
    \ev{\phi}\!(r,t) &\equiv \frac{1}{2} \ev{\phi}^\infty
                   \bigg[ 1 + {\tanh}\bigg(\frac{3\gamma_w(r-r_w(t))}{l_w}\bigg) \bigg] \,.
    \label{eq:wall-profile}
\end{align}
We denote the relativistic gamma factor associated with the wall velocity $\gamma_w$, and
\begin{align}
    r_w(t) &= r_w^0-v_w t \,
\end{align}
is the time-dependent position of the bubble wall in the plasma rest frame at time $t$. \Cref{eq:wall-profile} describes a spherical, contracting bubble wall reaching the origin after a time $r_w^0/v_w$.  In what follows, we will often shorten $m_\chi(r,t)$ to $m_\chi$, and we will use the notation $m_\chi^\infty$ to denote the mass of $\chi$ in the true vacuum.

Once we have numerically solved the Boltzmann equation to determine $\f$, we can calculate the number density and energy density of $\chi$ as a function of $r$ and $t$ by integrating over momenta. In momentum space, $p_r$ and $p_\sigma$ correspond to the axial and radial coordinates of a cylindrical coordinate system, so the number density and the energy density are
\begin{align}
    n_\chi(r,t)    &= g_\chi\int\!\frac{\upd p_r \, p_\sigma \upd p_\sigma}{(2\pi)^2} \f(r, p_r, p_\sigma,t) \,,\\
    \rho_\chi(r,t) &= g_\chi\int\!\frac{\upd p_r \, p_\sigma \upd p_\sigma}{(2\pi)^2} \,
                          [m_\chi^2(r,t) + p_r^2 + p_\sigma^2]^{1/2} \, \f(r, p_r, p_\sigma,t) \,.
    \label{eq:rho-chi-integral}
\end{align}
The energy density $\rho_\chi$ is the relevant quantity for BH formation, and we make use of the number density $n_\chi$ to validate our numerical results, as described in \cref{sec:continuity}.

\subsection{The Liouville Operator}
\label{sec:liouville}

The Liouville operator is the total time derivative of the phase space distribution function $\f(r, p_r, p_\sigma, t)$,
\begin{align}
	\boldsymbol{\mathrm L}[\f] = \td{\f}{t}
	&= \td{r}{t} \pd{\f}{r} + \td{p_r}{t} \pd{\f}{p_r} + \td{p_\sigma}{t} \pd{\f}{p_\sigma}
	 + \pd{\f}{t} \,.
    \label{eq:boltzmann-1}
\end{align}
Since we are interested in the scenario where $l_w \ll r_w(t)$, the Liouville operator can be approximated in different ways near the wall and in the bulk of the bubble.  We will first discuss the near-wall regime, before turning to the bulk.

\subsubsection{Near-Wall Regime}

Near the wall the $\chi$ particles have a non-negligible mass which can vary significantly with $r$ and $t$, and this provides the dominant contribution to the Liouville operator.  Since $l_w \ll r_w(t)$, we can treat our spatial coordinate system in the near-wall regime as quasi-Cartesian. Moreover, since most particles will spend very little time near the wall, we also neglect gravitational effects and collisions in this regime.

The radial velocity $\tdtext{r}{t}$ in \cref{eq:boltzmann-1} can be written as $p_r / E$, and the forces $\tdtext{p_r}{t}$ and $\tdtext{p_\sigma}{t}$ which act on the particle can also be rewritten as functions of position and momentum space coordinates. To do so, we start with the relativistic Lagrangian of a non-interacting massive particle,
\begin{align}
	\mathcal{L}_\text{free} = - \frac{m_\chi}{\gamma} \,,
\end{align}
with the Lorentz boost factor (in spherical coordinates)
\begin{align}
	\gamma = (1 - \dot r^2 - r^2{\sin^2}\theta \,\dot \varphi^2-r^2 \dot\theta^2)^{-1/2} \,.
	\label{eq:lorentz-factor}
\end{align}
We can then use the relations
\begin{align}
	E          &= \gamma \, m_\chi \,,                   &\qquad
	p_r        &= E \, \dot r \,,                        &\qquad
	p_\theta   &= E \,r \,\dot \theta \,,                &\qquad
	p_\varphi  &= E \,r  \sin{\theta} \, \dot\varphi \,, &\qquad
	p_\sigma^2 &\equiv p_{\theta}^2 + p_{\varphi}^2 \,,
	\label{eq:energy-momentum-definitions}
\end{align}
together with the Euler--Lagrange equations\footnote{Note that $p_r$, $p_\theta$, $p_\varphi$, and $p_\sigma$ denote physical momenta, not the canonical momenta in the Euler--Lagrange formalism.} to give
\begin{align}
    \td{p_r}{t} &= - \frac{m_\chi}{E} \pd{m_\chi}{r} + \frac{1}{r} \frac{p_\sigma^2}{E} \,,
                                                \label{eq:dprdt} \\
    \td{p_\sigma}{t}
        &= \td{p_\varphi}{t} \frac{p_\varphi}{p_\sigma} + \td{p_\theta}{t} \frac{p_\theta}{p_\sigma}
        = -\frac{1}{r} \frac{p_\sigma p_r}{E}\,.
    \label{eq:dpsigmadt}
\end{align}
The term $\propto \pdtext{m_\chi}{r}$ in the first line changes a particle's radial momentum $p_r$ as it interacts with the bubble wall.  The terms $\propto p_\sigma$ represent a Coriolis-type force, encoding the rotation of the momentum-space coordinate system as a $\chi$ particle moves non-radially.  The Liouville operator then takes the form
\begin{align}
	\boldsymbol{\mathrm L}[\f] &=
	    \frac{p_r}{E}\pd{\f}{r}
	  + \bb{ \frac{1}{r}\frac{p_\sigma^2}{E} - \frac{m_\chi}{E}\pd{m_\chi}{r}
	   } \pd{\f}{p_r}
	  - \frac{1}{r} \frac{p_\sigma p_r}{E} \pd{\f}{p_\sigma}
	  + \pd{\f}{t} \,.
    \label{eq:liouville-1}
\end{align}
Close to the wall the Coriolis-type terms $\propto 1/r$ are very small compared to the terms $\propto \pdtext{m_\chi}{r}\propto 1/l_w$.\footnote{While this may not be true for particles with very large tangential momenta, $p_\sigma \gg m_\chi^\infty$, this population is heavily Boltzmann suppressed.}  As such, we can drop these terms so that
\begin{align}
	\boldsymbol{\mathrm L}[\f] &=
	    \frac{p_r}{E}\pd{\f}{r}
	   - \frac{m_\chi}{E}\pd{m_\chi}{r} \pd{\f}{p_r}
	  + \pd{\f}{t} \,.
    \label{eq:liouville-2}
\end{align}
Finally, we insert our ansatz for $\f$, \cref{eq:f-ansatz}, to find
\begin{align}
    \boldsymbol{\mathrm L}[\f] &=
    \feq \, \bigg[
        \frac{p_r}{E} \pd{}{r}
        - \frac{m_\chi}{E} \pd{m_\chi}{r} \pd{}{p_r}
      + \pd{}{t} 
      - (1-\feq) \frac{m_\chi}{T_n E}v_w\pd{m_\chi}{r} 
                 \bigg] \A(r,p_r,p_\sigma,t) \,,
    \label{eq:liouville-3}
\end{align}
where we have used $\pdtext{m_\chi}{t} = v_w \pdtext{m_\chi}{r}$.

\subsubsection{Bulk Regime}

In the bulk of the bubble a reparameterisation of the momentum components is useful. Deep inside the bubble, we can assume $m_\chi \approx 0$. Furthermore, barring hard $\chi$--$\chi$ collisions, the energy $E$ of a $\chi$ particle is almost constant in time, with only minor changes caused by the gravitational term.  This motivates trading $p_r$ and $p_\sigma$ for
\begin{align}
	E   = \sqrt{p_r^2+p_\sigma^2}\,
	\qquad\text{and}\qquad
	\xi \equiv {\arctan} \ba{ \frac{p_r}{p_\sigma} } \,.
	\label{eq:p-xi}
\end{align}
The Liouville operator then becomes
\begin{align}
	\boldsymbol{\mathrm L}[\f] 
	&= \td{r}{t} \pd{\f}{r} + \td{E}{t} \pd{\f}{E} + \td{\xi}{t} \pd{\f}{\xi}
	 + \pd{\f}{t} \,.
    \label{eq:liouville-4}
\end{align}
The prefactor $\tdtext{r}{t}$ can simply be written as $\sin \xi$.  However, we need to take gravitational effects into account to determine $\tdtext{E}{t}$ and $\tdtext{\xi}{t}$.  Ignoring particle scattering and annihilation, which are accounted for by the collision term, a spherically symmetric gravitational field changes the energy of a $\chi$ particle moving from radius $r$ to radius $r'$ by a factor
\begin{align}
    \frac{E'}{E}=\sqrt{\frac{1-\frac{r_s}{r}}{1-\frac{r_s}{r'}}}\,,
    \label{eq:Eprime-over-E}
\end{align}
where the Schwarzschild radius $r_s$ is given by \cref{eq:rs}.\footnote{For the purpose of calculating the gravitational potential, we make the simplifying assumption that all mass contained in the false-vacuum pocket is concentrated at its centre. This introduces an $\mathcal{O}(1)$ error in our treatment of the gravitational force. However we will see below in \cref{fig:trajectory} that gravity is overall a subdominant effect, hence we refrain from implementing it in a more sophisticated way such as by using the interior rather than exterior Schwarzschild metric.} Considering an infinitesimal shift $r'=r+\upd r$ and $E'=E+\upd E$, we obtain
\begin{align}
    \td{E}{t} &= -\frac{r_s E \sin\xi}{2r(r-r_s)} \,.
    \label{eq:dpdt-grav}
\end{align}
To determine $\tdtext{\xi}{t}$, we use the gravitational deflection of a massless particle due to a spherically symmetric gravitational field,
\begin{align}
    \td{r}{\varphi}=\pm r^2{\sqrt{\frac{1}{b^2}-\frac{1}{r^2}\bigg(1-\frac{r_s}{r}\bigg)}}\,,
\end{align}
where $\varphi$ is the polar angle of $\chi$ with respect to the bubble centre (setting the azimuth angle $\theta=\pi/2$ without loss of generality) and the relativistic impact parameter is
\begin{align}
    b=\frac{L}{E}\frac{1}{\sqrt{1-\frac{r_s}{r}}}=\frac{r\cos \xi}{\sqrt{1-\frac{r_s}{r}}}\,.
\end{align}
Using the relations in \cref{eq:energy-momentum-definitions} and setting $\theta=\pi/2$, we can write $\tdtext{r}{\varphi} = r \dot{r} / r \dot{\varphi} = r p_r / p_\sigma = r \tan\xi$.  We then take the time derivative to find
\begin{align}
    \td{\xi}{t} &= \frac{|\sin\xi|}{2r^2}\frac{2r^3-r_s b^2}{r^3+r_s b^2}\ba{\frac{1}{b^2}-\frac{r-r_s}{r^3}}^{\!-1/2} \,.
    \label{eq:dxidt-grav}
\end{align}

Putting everything together, the Liouville operator in the bulk of the bubble becomes\footnote{In our implementation, we work with logarithmic reparametrisations of the $r$ and $t$ coordinates in the bulk. This is required to maintain numerical accuracy in late stages of the simulation, where $r_w(t)$ may be $\ll r_w^0$.}
\begin{align}
	\boldsymbol{\mathrm L}[\f] &= \feq \, \bigg[
	    \sin \xi \pd{}{r}
	  - \frac{r_s E \sin\xi}{2r(r-r_s)}  \pd{}{E}
	  + \frac{|\sin\xi|}{2r^2}\frac{2r^3-r_s b^2}{r^3+r_s b^2}
	    \ba{\frac{1}{b^2} - \frac{r-r_s}{r^3}}^{\!-1/2} \pd{}{\xi}
	                                                        \nonumber\\
	&\hspace{4 cm}
	  + \pd{}{t} +\frac{1-\feq}{T_n}\frac{r_s E \sin\xi}{2r(r-r_s)}\bigg]
	\A(r,E,\xi, t) \,.
	\label{eq:liouville-bulk}
\end{align}

\subsection{The Collision Term}
\label{sec:collision}

The right-hand side of the Boltzmann equation, \cref{eq:boltzmann}, describes the effect of $\chi \bar\chi \to \phi$ and $\chi \bar\chi \to \phi\phi$ annihilation, as well as scattering of $\chi$ particles with each other and with the scalar $\phi$. In our simulation we only include the collision term in the bulk of the bubble, where $m_\chi\approx 0$, since its impact near the wall of width $l_w\ll r_w$ is negligible. We now discuss the contributions to the collision term in the bulk one by one.

\subsubsection{The Inverse Decay Term Due to $\chi\bar\chi \leftrightarrow \phi$}

The general collision term for the process $\chi(p) + \bar\chi(q) \leftrightarrow \phi(k)$ is~\cite{Gondolo:1990dk}
\begin{align}
    \boldsymbol{\mathrm C}[f_\chi] &\supset
      - \frac{1}{2}\frac{1}{2E_p} \sum_\text{spins} \int \upd\Pi_q \upd\Pi_k  \,
        (2\pi)^4 \delta^{(4)}(p+q-k) \, |\mathcal M|^2 \nonumber\\
    &\phantom{\supset} \times
        \Big[ f_{\chi,p} f_{\bar\chi,q} (1 + f_{\phi,k}) 
            - f_{\phi,k} (1 - f_{\chi,p}) (1 - f_{\bar\chi,q}) \Big] \,,
\end{align}
where $\mathcal M$ is the matrix element; $p$, $q$ and $k$ are the momenta of the incoming and outgoing particles with corresponding energies $E_k = ( m_\phi^2 + |\vec k|^2 )^{1/2}$ (and analogously for $E_p$ and $E_q$); and the integration measure for each 3-momentum integral is $\upd\Pi_k = \upd^3k / (16 \pi^3 E_k)$. For the phase space distribution functions, we have introduced the short-hand notation $f_{\chi,p} \equiv f_\chi(r, p_r, p_\sigma, t)$ and $f_{\phi,k} \equiv f_\phi^\text{eq}(\vec k)$.\footnote{While in our code we ultimately express the collision term in terms of $E$ and $\xi$ according to \cref{eq:p-xi}, we use the more intuitive momentum variables $p_r$ and $p_\sigma$ in the following discussion.} The prefactor $1/2$ comes about because \f stands for the phase space distribution function of a single degree of freedom of $\chi$.

Even with the large deviations from equilibrium that will arise in our calculations, it is safe to neglect Bose enhancement and Pauli blocking by setting $1 \pm f = 1$.\footnote{The overdensities that build up due to reflections off the wall are localised around large momenta $\sim m_\chi^\infty$. For this reason, we observe $\feq \ll \f$ but $\f \ll 1$.} Using
\begin{align}
    \sum_\text{spins}|\mathcal M|^2 = 4y_\chi^2(p\cdot q)\,,
\end{align}
we arrive at 
\begin{align}
    \boldsymbol{\mathrm C}[f_\chi] &\supset
        - \frac{y_\chi^2}{E_p} \int \upd\Pi_q \upd\Pi_k  \,
        (2\pi)^4 \delta^{(4)}(p+q-k) \, (p\cdot q) \big[ f_{\chi,p}f_{\bar\chi,q}  - f_{\phi,k} \big] \\
        &=- \frac{y_\chi^2\,m^2_\phi}{32\pi^2} \frac{f^\text{eq}_{\chi,p}}{E_{p}}\int\upd \alpha_{pq} \upd q_r \upd q_\sigma \frac{q_\sigma}{E_{q}(m_\phi^2+|\vec p+\vec q|^2)^{1/2}} \,
        \delta(E_{p}+E_{q}-(m_\phi^2+|\vec p+\vec q|^2)^{1/2}) \\
        &\qquad\times \big[ \A(r,p_r,p_\sigma,t) \A(r,q_r,q_\sigma,t) - 1 \big]f^\text{eq}_{\bar\chi,q}\nonumber\\
        &=- \frac{y_\chi^2\,m^2_\phi}{32\pi^2}\frac{f^\text{eq}_{\chi,p}}{E_{p}\,p_\sigma} \int \frac{\upd q_r \upd q_\sigma}{E_{q}}\Bigg[1-\ba{\frac{2E_{p} E_{q}-2p_rq_r-m_\phi^2}{2 p_\sigma q_\sigma}}^{\!2}\Bigg]^{-1/2} \nonumber\\
\label{eq:collision-1}
        &\qquad\times\big[ \A(r,p_r,p_\sigma,t) \A(r,q_r,q_\sigma,t) - 1 \big]f^\text{eq}_{\bar\chi,q}\,,
\end{align}
where $\alpha_{pq}$ is the angle between the tangential components of $\vec p$ and $\vec q$ (or, alternatively, the angle between the planes containing the trajectories of $\chi(p)$ and $\bar\chi(q)$). To integrate over $\vec k$, we have made use of detailed balance, $f_{\phi,k}\approx f^\text{eq}_{\chi,p}f^\text{eq}_{\bar\chi,q}$. The integral over $\alpha_{pq}$ implies the kinematic requirement $|2E_{p}E_{q}-2p_rq_r-m_\phi^2|<2p_\sigma q_\sigma$.

In the scenario we are considering, where the $\chi$ particles are compressed by the bubble walls, the phase space enhancement factor \A is always larger than or equal to one. The collision term, \cref{eq:collision-1}, is then always negative, corresponding to annihilation (rather than production) of $\chi \bar\chi$. Therefore, this contribution to the collision term counteracts $\chi$ build-up due to compression.

\subsubsection{The Annihilation Term Due to $\chi\bar\chi \leftrightarrow \phi\phi$}

The $\chi(p) + \bar\chi(q) \leftrightarrow \phi(k) + \phi(l)$ contribution to the collision term is given by~\cite{Gondolo:1990dk}
\begin{align}
    \boldsymbol{\mathrm C}[f_\chi] &\supset
      - \frac{1}{2}\frac{1}{2E_p} \sum_\text{spins} \int \upd\Pi_q \upd\Pi_k \upd\Pi_l \,
        (2\pi)^4 \delta^{(4)}(p+q-k-l) \, |\mathcal M|^2 \nonumber\\
    &\phantom{\supset} \times
        \Big[ f_{\chi,p} f_{\bar\chi,q} (1 + f_{\phi,k}) (1 + f_{\phi,l})
            - f_{\phi,k} f_{\phi,l} (1 - f_{\chi,p}) (1 - f_{\bar\chi,q}) \Big] \,.
    \label{eq:annihilation-term-1}
\end{align}
After again setting $1\pm f=1$ and making use of detailed balance, $f_{\phi,k} f_{\phi,l} \approx f^\text{eq}_{\chi,p} f^\text{eq}_{\bar\chi,q}$, we can integrate \cref{eq:annihilation-term-1} over $k$ and $l$ to arrive at
\begin{align}
    \boldsymbol{\mathrm C}[f_\chi] &\supset
      - \frac{f_{\chi,p}^\text{eq}}{E_p}
        \int \upd\Pi_q \, 4 F\, \sigma(\chi\bar\chi\to\phi\phi) \,
        \big[ \A(r, p_r, p_\sigma, t) \A(r, q_r, q_\sigma, t) - 1 \big] f_{\bar\chi,q}^\text{eq} \,,
    \label{eq:annihilation-term-2}
\end{align}
where the kinematic factor,
\begin{align}
    F \equiv E_p E_q |v_\chi - v_{\bar\chi}|
      =      \frac{1}{2} \sqrt{(s-2m_\chi^2)^2 - 4 m_\chi^4} \,,
\end{align}
comes from rewriting the squared matrix element in terms of the cross section $\sigma(\chi\bar\chi\to\phi\phi)$ \cite{Gondolo:1990dk}. Note that an extra factor of two arises as the cross section is defined in terms of an average over both initial spin states, whereas the Boltzmann equation requires an average over only one of them, see \cref{eq:annihilation-term-1}.

In the bulk of the bubble, where $m_\chi \approx 0$, the annihilation cross section is
\begin{align}
    \sigma(\chi\bar\chi \to \phi\phi) =
        \frac{y_\chi^4}{32 \pi s} \big[ 2\log(s/m_\phi^2) - 3 \big]
      + \mathcal O(m_\phi^2/s) \,,
    \label{eq:annihilation-xsec}
\end{align}
with the squared centre-of-mass energy
\begin{align}
    s = 2 (E_p E_q - p_r q_r - p_\sigma q_\sigma \cos\alpha_{pq}) \,.
    \label{eq:s-mchi0}
\end{align}
Neglecting terms suppressed by powers of $m_\phi^2/s$ is justified in \cref{eq:annihilation-xsec} because the overdensities appear at energies $\gg T_n \sim m_\phi$.  Inserting \cref{eq:annihilation-xsec,eq:s-mchi0} into \cref{eq:annihilation-term-2} and integrating over $\alpha_{pq}$,\footnote{Integrating over $\alpha_{pq}$ before integrating over the other momenta is, strictly speaking, incorrect because the integration limits for these momenta in \cref{eq:annihilation-term-2} depend on $\alpha_{pq}$ (that is, the integral should only extend over the region $s>4m_\phi^2$). However, we ignore this and instead approximate the integration region in \cref{eq:annihilation-term-3} to be $E_p E_q - p_r q_r - p_\sigma q_\sigma > 2 m_\phi^2$.} we arrive at
\begin{align}
    \boldsymbol{\mathrm C}[f_\chi] &\supset
      - \frac{y_\chi^4}{128\pi^3}  \frac{f_{\chi,p}^\text{eq}}{E_{p}}
        \int \! \upd q_r \upd q_\sigma \, \frac{q_\sigma}{E_q}
        \big[ \A(r,p_r,p_\sigma,t) \A(r,q_r,q_\sigma,t) - 1 \big] f_{\bar\chi,q}^\text{eq} \nonumber\\
    &\phantom{\supset} \times
      \bb{ 2\,{\log} \ba{\frac{p_\sigma q_\sigma}{m_\phi^2}}
         + 4\,{\text{arsinh}}\ba{ \sqrt{\frac{E_p E_q-p_rq_r-p_\sigma q_\sigma}{2 p_\sigma q_\sigma}} } - 3} \,.
    \label{eq:annihilation-term-3}
\end{align}
This contribution to the collision term also counteracts $\chi$ build-up due to compression.

\subsubsection{The Momentum Redistribution Terms Due to $\chi$ Scattering}

While the annihilation terms discussed above reduce the $\chi$ number density, the scattering processes $\chi\chi \leftrightarrow \chi\chi$, $\chi\bar\chi \leftrightarrow \chi\bar\chi$ and $\chi\phi \leftrightarrow \chi\phi$ conserve the total number of $\chi+\bar\chi$. The only effect of these processes is the redistribution of phase space density from large momenta $\sim m_\chi^\infty$, where overdensities build up due to the reflections off the wall, towards smaller momenta. This boosts annihilation, which is most efficient for small energies, but at the same time reduces the amount of particles that enter the true vacuum by passing through the wall. However, we will see in \cref{sec:results} that parameter points that lead to successful black hole formation are those for which the $\chi\bar\chi$ annihilation rate is small. Since the 2-to-2 annihilation and scattering rates scale in the same way as functions of the model parameters, this implies that momentum redistribution will also be negligible at the parameter points of most interest to us.

\subsection{Numerically Solving the Boltzmann Equation}
\label{sec:implementation}

Putting together the Liouville operator and the collision term, the Boltzmann equation in the near-wall regime takes the form
\begin{align}
  \Omega_r\pd{\A}{r} + \Omega_{p_r}\pd{\A}{p_r} + \pd{\A}{t} = \Omega_\A\,,
  \label{eq:final-boltzmann}
\end{align}
where
\begin{align}
 \Omega_r(r,p_r, p_\sigma,t)     &\equiv   \frac{p_r}{E} \,,\\
 \Omega_{p_r}(r,p_r, p_\sigma,t) &\equiv -\frac{m_\chi}{E} \pd{m_\chi}{r} \,,\\
 \Omega_\A(r,p_r, p_\sigma,t,\A) &\equiv
    (1-\feq) \bigg(\frac{m_\chi}{T_n E}v_w\pd{m_\chi}{r} \bigg) \A \,.
\end{align}
\Cref{eq:final-boltzmann} is a  partial differential equation (PDE) for the function $\A(r,p_r,p_\sigma,t)$.  Since the $\Omega$ coefficients on the left-hand side of \cref{eq:final-boltzmann} are independent of $\A$, and since $\Omega_{\A}$ does not contain any derivatives of $\A$, the \textit{method of characteristics} can be used to reduce the PDE to an infinite set of ordinary differential equations (ODEs) that are coupled only through the collision term $\boldsymbol{\mathrm C}[f_\chi]$ (see ref.~\cite{Baker:2019ndr} for an analogous application). This method involves first solving the coupled ODEs
\begin{align}
    \td{r(t)}{t} = \Omega_r \,,
    \qquad
    \td{p_r(t)}{t} = \Omega_{p_r} \,,
    \qquad
    \td{p_\sigma(t)}{t} = 0 \,,
    \label{eq:lagrange-charpit}
\end{align}
for a large set of different initial conditions, to construct a set of curves in $r$--$p_r$--$p_\sigma$ space. Each curve $(r(t), p_r(t), p_\sigma(t))$ can be interpreted as a particle trajectory in the absence of the collision term (which we set to be zero in the near-wall regime in any case).  Curves incident on the wall with a momentum much smaller than the $\chi$ mass gain are reflected while curves with much larger momenta will pass through the wall.  The solution for $\A(t)$ along each curve can then be found by solving the ODE
\begin{align}
    \label{eq:ode-1}
    \td{\A(t)}{t} = \Omega_\A\big(r(t),p_r(t), p_\sigma(t),t,\A(t) \big)\,.
\end{align}

When a curve leaves the near-wall regime and enters the bulk, which we take to occur when $r(t) < r_w(t)-3 l_w$, we switch to the Boltzmann equation relevant for the bulk.  In the bulk the Boltzmann equation takes the form
\begin{align}
  \tilde\Omega_{r}\pd{\A}{ r} + \tilde\Omega_{E}\pd{\A}{E} + \tilde\Omega_{\xi}\pd{\A}{\xi} 
  + \pd{\A}{t}
        = \tilde\Omega_\A\,,
  \label{eq:final-boltzmann-bulk}
\end{align}
where the $\tilde\Omega$'s can be determined from \cref{eq:liouville-bulk,eq:collision-1,eq:annihilation-term-3}.  We again use the method of characteristics to solve this equation.  Using the left-hand side of \cref{eq:final-boltzmann-bulk} we determine a large set of particle trajectories in the absence of the collision term.  We then solve
\begin{align}
    \label{eq:ode-2}
    \td{\A(t)}{t} = \tilde\Omega_\A \big(r(t), E(t), \xi(t), t,\A(t) \big)\,,
\end{align}
using a logarithmic reparametrisation of $r$ and $t$, to find $\A(t)$ along each curve.

To evolve \cref{eq:ode-1,eq:ode-2} in time, we start with the initial condition $\A(t_0) = 1$ for all $r$, $p_r$ and $p_\sigma$.  The values of $\A$ along these curves can then be used to determine the full $\A(r,p_r, p_\sigma,t)$.  In practice, we do this by discretising the $r$--$p_r$--$p_\sigma$--$t$ and $r$--$E$--$\xi$--$t$ spaces and averaging over all curve segments which pass through a given grid cell.  When a curve passes from the near-wall to the bulk regime (or vice versa), we use the final $\A$ in the previous regime as an initial condition in the new regime.

Note that in the annihilation terms the interaction partners are $\chi$ or $\bar\chi$, so have the same phase space distribution as the primary particle. The phase space integrals in \cref{eq:collision-1,eq:annihilation-term-3} therefore couple the ODEs in \cref{eq:ode-2} through the $\chi$ momenta.  To deal with this, we evolve all ODEs simultaneously in time, using $\A(r,q_r,q_\sigma,t-\Delta t)$ as an approximation for the phase space enhancement of the interaction partner in the integrals of the collision term at time $t$.

There are four different types of trajectories that interact with the bubble wall:
\begin{enumerate}[label=(\Alph*)]
    \item Particles starting inside the shrinking bubble, with initial $p_r \ll m_\chi^\infty$. These particles will be reflected by the wall, leading to an energy overdensity.  At some point this may be large enough for a black hole to form, at which point these particles are lost inside.
    \item Particles starting inside the shrinking bubble, with initial $p_r \lesssim m_\chi^\infty$. Since the $\chi$ particles gain momentum on every reflection, some particles will eventually have enough momentum to pass through the bubble wall.\footnote{The momentum threshold below which particles are reflected is not exactly $m_\chi^\infty$, but smaller due to the wall's motion.  For $p_\sigma = 0$, for instance, the threshold occurs at $p_r = m_\chi^\infty/[\gamma_w(1+v_w)]$. It would be exactly $m_\chi^\infty$ in the wall's rest frame.}  They then leave towards $r\to\infty$. As they now move relatively slowly, since they are massive, they could still end up inside the black hole once it forms.  However, we conservatively neglect this contribution in our simulations.
    \item Particles starting inside the bubble, with initial $p_r \gtrsim m_\chi^\infty$. When these particles reach the bubble wall, they will be able to pass through it while gaining mass and losing momentum. They then leave towards $r\to\infty$.  When $m_\chi^\infty \gg T_n$, there will be almost no particles of this type since the high energy tails of the equilibrium distribution are exponentially suppressed. Particles of this type therefore do not play a significant role for black hole formation.
    \item Particles starting with mass $m_\chi^\infty$ outside the shrinking bubble in the $\ev{\phi} \neq 0$ phase and moving towards the bubble wall. If they reach the wall before the bubble has collapsed, they will enter the bubble, losing their mass and gaining an equivalent amount of kinetic energy. From then on, they belong to category (C).  When $m_\chi^\infty \gg T_n$, there will again be almost no particles of this type due to Boltzmann suppression of their initial abundance.
\end{enumerate}
We include categories (A) and (B) in our simulation, although trajectories of type (B) are only tracked until they pass through the bubble wall.  The repeated reflection of these particles off the wall results in a gradual increase of the phase space enhancement factor $\A$ and the energy density inside the shrinking bubble. 

\begin{figure}
    \centering
    \includegraphics{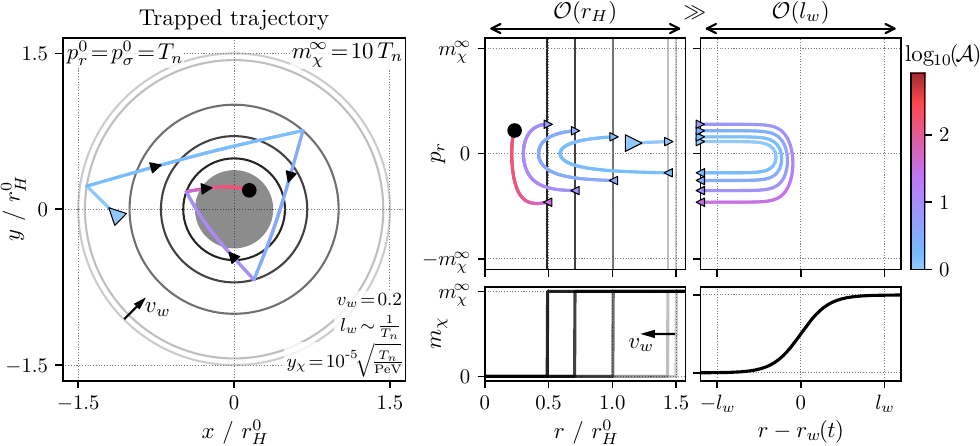}\\[1em]
    \hspace{3.65cm} (i) \hspace*{\fill} (ii) \hspace{4.51cm}
    
    \vspace{0.8cm}
    \includegraphics{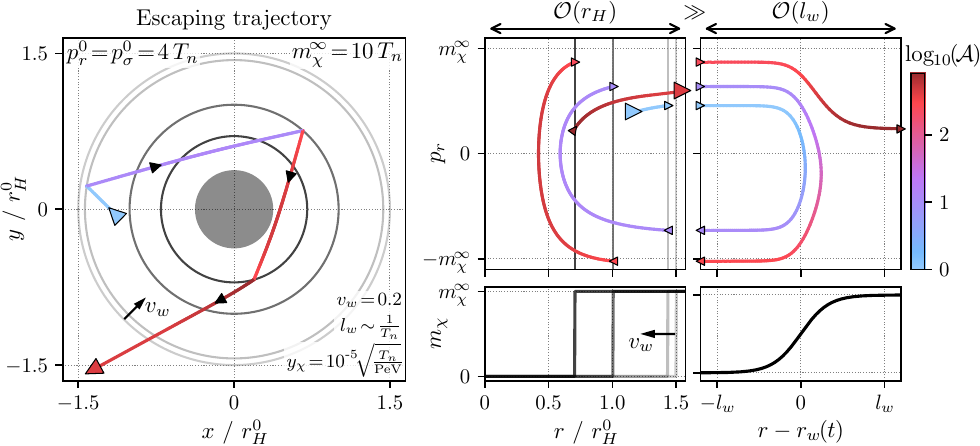}\\[1em]
    \hspace{3.575cm} (iii) \hspace*{\fill} (iv) \hspace{4.49cm}
    \caption{Two example trajectories in position space and phase space.  Panels (i) and (ii) both show the same type (A) trajectory, which is reflected by the bubble wall several times and remains inside the bubble until a black hole forms (grey disc).  Panels (iii) and (iv) show a type (B) trajectory, which passes through the bubble wall before the black hole forms.  We show the trajectories in position space on the left, in (i) and (iii), while on the right, in (ii) and (iv), we illustrate the same trajectories in $r$--$p_r$ phase space.  The top two sub-panels in (ii) and (iv) show the $r$--$p_r$ plane far away from (left) and near to (right) the wall. The small coloured arrowheads show where the trajectories pass between the bulk regime and the near-wall regime.  In the lower sub-panels of (ii) and (iv) we show the wall profile at the time of each reflection.  The colour of the trajectories represents the phase space enhancement factor $\A$.  The starting position of the trajectories is indicated with a large blue arrowhead, while a black dot or a large red arrowhead indicates the moment of black hole formation or when the trajectory leaves the simulation, respectively.
    }
    \label{fig:trajectory}
\end{figure}

Two example trajectories are shown in \cref{fig:trajectory}.  On the left we show trajectories in position space, and on the right we show the same trajectories in the $r$--$p_r$ phase space.  When in equilibrium, at the start of the simulation, most of the $\chi$ particles have $p_r\sim p_\sigma \sim T_n$, so the top row shows the evolution for a representative region of phase space. In panel (i) we see a trajectory that starts on the left, just inside the bubble wall, and moves in the $(-1,1)$ direction in the $(x,y)$ plane before being reflected several times by the bubble wall.  The position of the bubble wall at the time of each reflection is indicated by black circles.  The colour of the trajectory indicates the enhancement factor $\A$ at that position and time.  We see that $\A$ increases as the trajectory spirals in, until a black hole finally forms.  For the parameters chosen, there is negligible annihilation, so $\A$ predominantly increases with time.  Shortly before the black hole forms, we can see the bending of the trajectory due to gravity. The fact that the trajectory bends only slightly confirms the simplified and approximate treatment of gravity in \cref{eq:Eprime-over-E}. Note that this trajectory is of type (A).  

In \cref{fig:trajectory} (ii) we show four sub-panels.  The top row shows the trajectory from (i) in $r$--$p_r$ phase space, and the lower row shows the bubble wall profile at each reflection.  In the left sub-panels we show the trajectory inside the bulk of the bubble, while on the right we zoom in on its interaction with the bubble wall (notice that the right sub-panel is plotted as a function of $r-r_w(t)$; the wall does not move in this frame).  We see that the overdensity $\A$ slowly increases, and that on each reflection the trajectory gains radial momentum.  Although the momentum is increasing, it does not become large enough to escape the bubble by the time the black hole forms.

For the bottom row of \cref{fig:trajectory}, we have chosen a trajectory with a larger initial momentum, $p_r^0 = p_\sigma^0 = 4\,T_n$, corresponding to a particle which reflects twice before gaining enough momentum to pass through the bubble wall.  This trajectory belongs to category (B).

For illustration we have chosen $m_\chi^\infty = 10\,T_n$, wall velocity $v_w = 0.2$, and wall thickness $l_w \sim 1/T_n$ in \cref{fig:trajectory}, as indicated in the plot. For the Yukawa coupling, we choose a value
\begin{align}
    y_\chi = \num{e-5} \sqrt{\frac{T_n}{\si{PeV}}} \,.
    \label{eq:ychi}
\end{align}
With this choice, \cref{fig:trajectory} (and all subsequent results, unless indicated otherwise) are essentially independent of $T_n$, as long as $T_n$ is not too high. The rate for the main annihilation process $\chi\bar\chi \to \phi$ scales as $y_\chi^2 T_n$, while $r_H^0$ scales as $1/T_n^2$ in a radiation-dominated Universe. \Cref{eq:ychi} then ensures that the number of annihilations per Hubble length $r_H^0$ remains invariant.  If $T_n$ is higher ($T_n \gtrsim \SI{e10}{TeV}$, for the value of $y_\chi$ chosen in \cref{fig:trajectory}), annihilation via $\bar\chi \chi \to \phi\phi$, whose rate scales as $y_\chi^4 T_n$, becomes important as well, and the results are no longer independent of $T_n$.

\subsection{Consistency Check}
\label{sec:continuity}

To ensure the validity of our numerical results, we checked that the continuity equation is fulfilled. We compute the total particle number trapped inside the compressing bubble as a function of time,
\begin{align}
  N_\chi(t)=4\pi\int_{0}^{r_w(t)}\upd r\,r^2 \,n_\chi(r,t) \,,
\end{align}
from our simulation results, and we check that any change of this quantity is accounted for by the particles escaping through the bubble wall and the particles lost to annihilation. In other words, we check that
\begin{align}
  N_\chi(t) - \int_{t_0}^{t} \upd t' \,\Big([\dot N_\chi(t')]_\text{escaping}
            + [\dot N_\chi(t')]_\text{annihilation}\Big)
    = N_\chi(t_0) \,,
    \label{eq:crosscheck}
\end{align}
with
\begin{align}
  &[\dot N_\chi(t)]_\text{escaping}
    = -g_\chi\bb{4\pi r^2 \int\frac{\upd^3 p}
                                   {(2\pi)^3}
                              \frac{p_r+v_w E}{E} \f}_{r=r_w(t)+3l_w} \,,
    \label{eq:N-escaping}
\intertext{and}
  &[\dot N_\chi(t)]_\text{annihilation}
    =  g_\chi\,4\pi\int_{0}^{r_w(t)} \upd r\,r^2
       \int\frac{\upd^3 p}
                                   {(2\pi)^3}
           \boldsymbol{\mathrm C}[\f] \,,
\end{align}
is satisfied. Note that, to obtain the rate at which particles escape, we evaluate the phase space distribution function at a distance of $3l_w$ from the wall centre in the true vacuum, where $\pdtext{m_\chi}{r}$ is approximately zero and reflections occurring beyond this point are negligible.  Also note that to obtain \cref{eq:N-escaping}, we compute $[\dot N_\chi(t)]_\text{escaping}$ in the wall's rest frame and then transform back to the plasma frame. This ensures that particles that are moving inwards ($p_r < 0$ in the plasma frame), but that have been overtaken by the wall, count as escaping, as they should.

We will present and discuss the results of this consistency check for several benchmark parameter points in \cref{sec:results}.

\section{Results}
\label{sec:results}

We are now ready to present our numerical results for the evolution of the $\chi$ population inside a shrinking bubble, and to demonstrate black hole formation. We will first illustrate the evolution of the $\chi$ phase space density during the phase transition, followed by results for the $\chi$ energy density for a number of illustrative benchmark points. While we cannot compute the properties of the resulting black hole population, such as the black hole mass and cosmological abundance, without an explicit potential, some general remarks can be found in ref.~\cite{Baker:2021nyl}.

\subsection{The $\chi$ Phase Space Distribution}
\label{sec:phase-space-distribution-function}

\begin{figure}
    \centering
    \hspace*{\fill}
    \begin{tabular}{c}
        {\large $r_w(t_1) \approx 1.5\,r_H^0$}\\[0.3em]
        \includegraphics{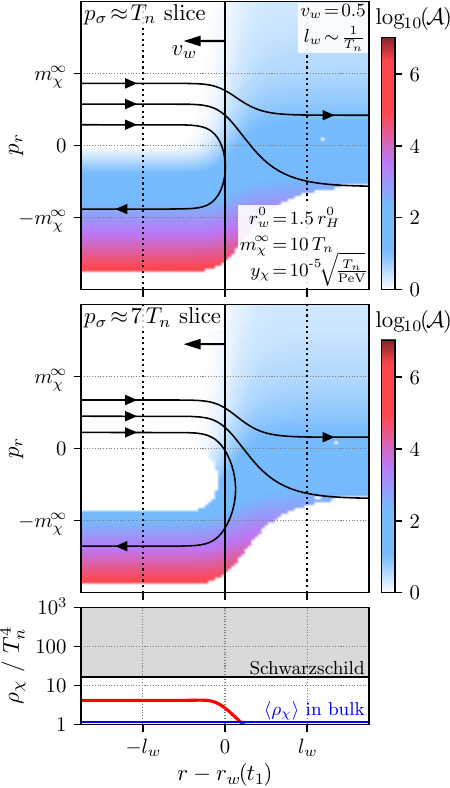}
    \end{tabular}
    \hfill
    \begin{tabular}{c}
        {\large $r_w(t_2) \approx 0.5\,r_H^0$}\\[0.3em]
        \includegraphics{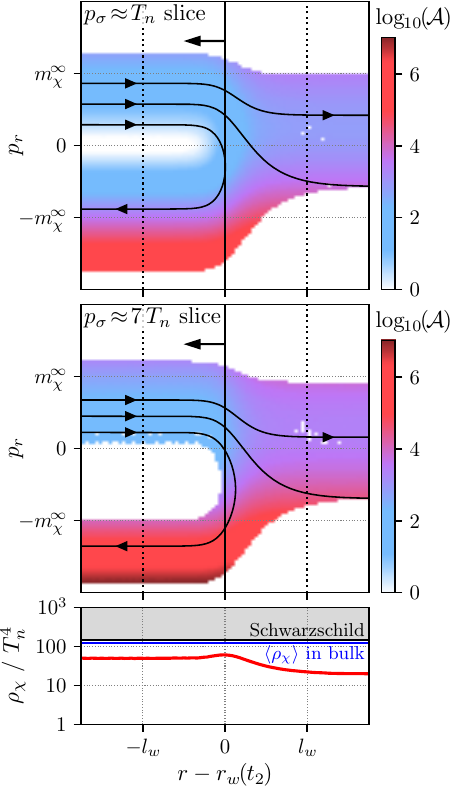}
    \end{tabular}
    \hspace*{\fill}
    \caption{The phase space enhancement factor $\A = \f / \f^\text{eq}$ in slices of fixed $p_\sigma$ and at a fixed time (upper and middle panels), and the energy density of $\chi$ at a fixed time (lower panels), as a function of radius $r$ near the wall. On the left, we show the initial situation shortly after the start of the simulation. An overdensity has already been generated in front of the wall, caused by reflected particles that were previously in equilibrium. On the right, we show the same scenario at a later time, after the radius of the bubble has shrunk by a factor of three, so that large overdensities have accumulated in front of the wall and the Schwarzschild criteria (grey region in the bottom panels) is almost met. The black contours in the top and middle panels depict illustrative phase space trajectories.}
    \label{fig:slices}
\end{figure}

In \cref{fig:slices} we show two snapshots of the phase space enhancement factor $\A(r,p_r,p_\sigma,t)$ in the vicinity of the wall, for the benchmark parameters $r_w^0 = 1.5 \,r_H^0$, $m_\chi^\infty = 10 \,T_n$, $y_\chi = 10^{-5} \sqrt{T_n/\text{PeV}}$, $v_w = 0.5$ and $l_w \sim 1/T_n$.  On the left we show $\A$ shortly after the beginning of the simulation, at $r_w(t_1) \approx 1.5\,r_H$.  On the right we show $\A$ at a later time when the bubble has shrunk by a factor of three, $r_w(t_2) \approx 0.5\,r_H$.  When in equilibrium, most $\chi$ particles have $p_r\sim p_\sigma \sim T_n$, so the top row shows the evolution for a representative $p_\sigma$, while the middle row shows a higher momentum slice with $p_\sigma \approx 7\,T_n$.  The bottom panels in \cref{fig:slices} show the $\chi$ energy density near the wall, the average $\chi$ energy density in the bulk of the bubble, \cref{eq:rho-chi-integral}, and the average energy density required to satisfy the Schwarzschild criteria for a bubble of radius $r_w(t)$.  The horizontal axis in all plots is in units of $r-r_w(t_i)$, so the centre of the wall is located at the origin and is moving to the left.

On the left of \cref{fig:slices}, which shows the phase space density shortly after the start of the simulation, we see for both $p_\sigma \approx T_n$ (top) and $p_\sigma \approx 7\,T_n$ (middle) that, by assumption, $\mathcal{A}=1$ for particles approaching the wall ($p_r > 0$, $r-r_w(t_1) \lesssim l_w$).  These particles are reflected from the wall leading to an overdensity, $\mathcal{A}>1$, in the $p_r < 0$, $r-r_w(t_1) \lesssim l_w$ region of phase space.  This increase in $\A$ as $\chi$ particles reflect from the wall is the key driver of the increasing $\chi$ energy density as the bubble shrinks, which will ultimately lead to black hole formation.  There is also a slight overdensity at $r - r_w(t_1) \gtrsim l_w$ due to $\chi$ particles which have fully traversed the wall. However, recalling that $\A$ is the \emph{relative} overdensity compared to $\f^\text{eq}$ and that $\f^\text{eq}$ is very small beyond the wall due to the large $\chi$ mass, far fewer particles traverse the wall than are reflected.  The black contours indicate that $|p_r|$ is larger after a reflection, showing the momentum boost a $\chi$ particle receives from the wall.  In the bottom-left panel we see, as expected, that the energy density is much larger than $\rho_\chi^\text{eq} \approx T_n^4$ inside the shrinking bubble ($r-r_w(t_1) < 0$), but much smaller than $T_n^4$ outside the bubble wall ($r-r_w(t_1) > 0$).  The average energy density in the bulk of the bubble is still close to its equilibrium value, $\rho_\chi^\text{eq} \sim T_n^4$, even though $\A$ is much larger than one near the wall, as the overdensity has not yet had time to fill the whole bubble.

The right side of \cref{fig:slices} shows a snapshot at a later time, where $r_w(t)$ has shrunk by a factor of three.  We see that there is now a large overabundance of particles approaching the wall.  This increase has occurred both because most particles are trapped inside the shrinking bubble, and because they receive a momentum boost on each reflection. The enhancement is more pronounced at $p_\sigma \approx 7\,T_n$ than at $p_\sigma \approx T_n$, as can be understood as follows.  Firstly, particles with larger $p_\sigma$ are reflected more frequently. Secondly, more energetic particles receive a larger boost on each reflection.  Therefore, at large $p_\sigma$, reflections populate phase space regions with smaller \feq than is the case for small $p_\sigma$, leading to a larger enhancement of $\A$.  We see, however, that the overdensity does not extend far beyond $p_r \approx m_\chi^\infty$ since particles with such large momenta are able to pass through the wall and escape the bubble.  We see from the scale of $\A$ that the overdensity inside the bubble is now very large, and that reflection again substantially increases the overdensity (and so the energy density inside the bubble).  Indeed, the bottom-right panel shows that $\ev{\rho_\chi} \sim 100\,T_n^4$ now, and that the black hole formation criteria is almost satisfied.\footnote{The reader may wonder why $\ev{\rho_\chi}$ in the bubble (horizontal blue line) is larger than $\rho_\chi(r)$ (red curve) at any $r$ near the wall in the bottom-right panel of \cref{fig:slices}. This is because some of the particles approaching the wall at this particular point in time have not yet reflected off the wall and are therefore still missing out on the associated $\A$ enhancement and momentum boost.  In the bulk, however, there is a large population of once-reflected particles travelling approximately radially that have not yet had time to cross the whole bubble, but which contribute to~$\ev{\rho_\chi}$.}  It also shows that, while a significant number of particles can now traverse the wall, their overall energy density is still small.  This is again because $\mathcal{A}$ is the overdensity relative to $\f^\text{eq}$, which is heavily suppressed outside the bubble. 

Overall we see that as the bubble shrinks, the majority of $\chi$ particles are reflected by the wall and a large energy overdensity builds up inside the shrinking region.

\subsection{The $\chi$ Energy Density}
\label{sec:energy-density}

\begin{figure}
    \centering
    \includegraphics{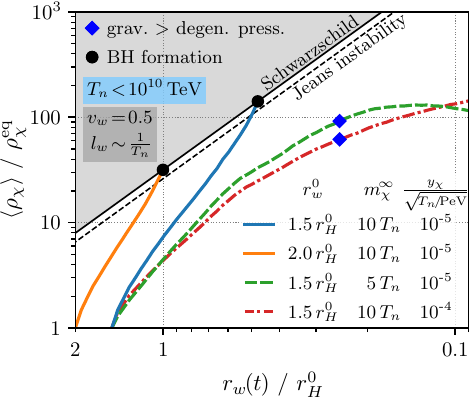}\hfill\includegraphics{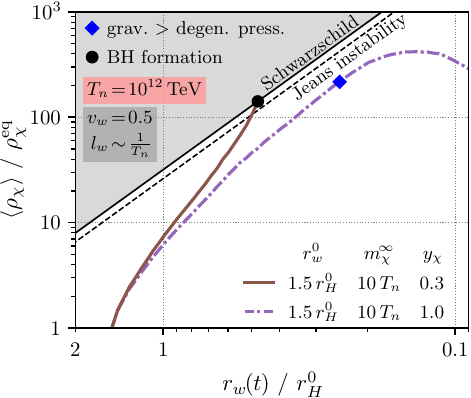}
    \caption{
        The $\chi$ energy density averaged over the bubble, $\ev{\rho_\chi}$, as a function of the bubble radius for several parameter points at $T_n\lesssim \SI{e10}{TeV}$ (left) and $T_n = \SI{e12}{TeV}$ (right).  The blue, orange, and brown curves lead to black hole formation once the Schwarzschild criteria is satisfied (grey region), while the dashed green, dot-dashed red, and dot-dashed purple curves do not.
    }
    \label{fig:rho}
\end{figure}

\begin{figure}
    \centering
    \includegraphics{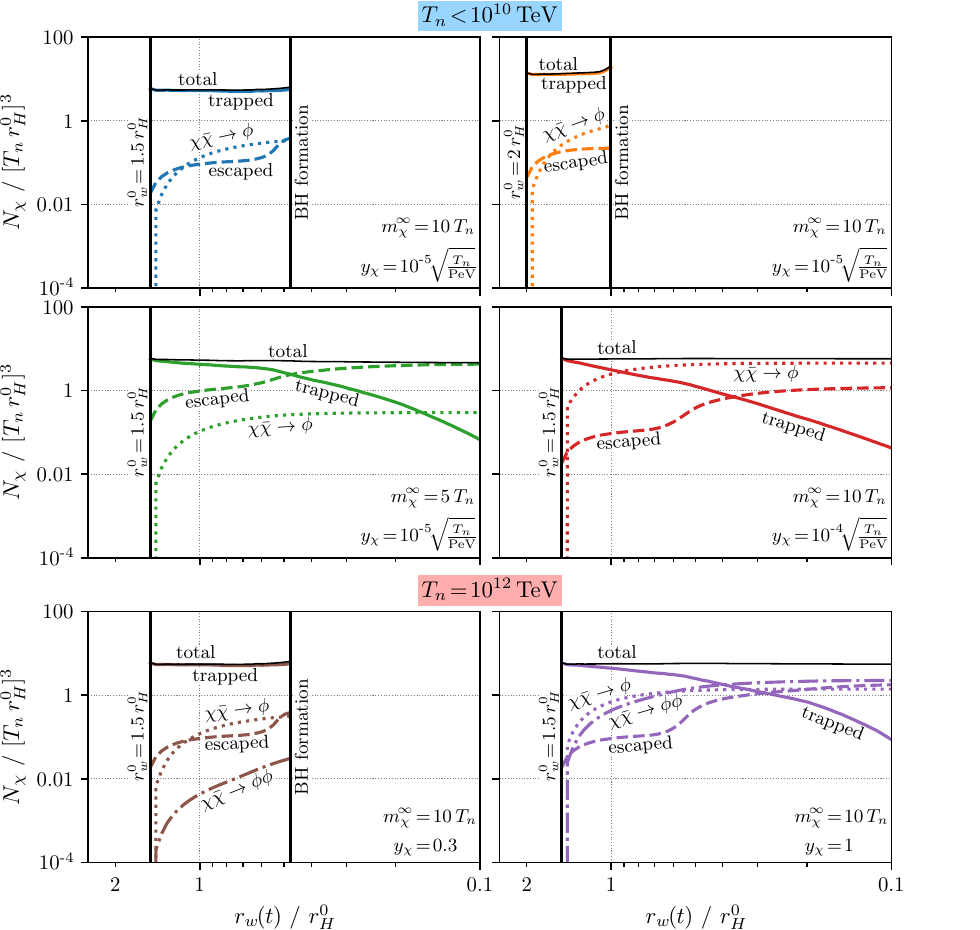}
    \caption{
        Particle number $N_\chi$ of $\chi$ inside the shrinking bubble as a function of the bubble radius $r_w(t)$ (solid), compared to the number of particles that have left the simulation before the wall has shrunk to a given radius by either entering the true vacuum (dashed), or by annihilation (dotted and dot-dashed). Each panel corresponds to a curve in \cref{fig:rho}. The sum of all coloured curves, shown as a thin black line, is almost constant in each panel, indicating that our simulation fulfils the continuity equation up to small numerical artefacts, mostly from discretisation.
        \label{fig:crosscheck}
    }
\end{figure}

Shifting our focus from the local phase space density in the vicinity of the wall to the global density in the whole bubble, we plot in \cref{fig:rho} the $\chi$ energy density averaged over the bubble as a function of time, parameterised in terms of the shrinking bubble's radius.  We again fix the wall velocity and wall thickness, deferring discussion of variations in these quantities to \cref{sec:dependence-wall-velocity}. In the left-hand panel, we plot in blue the curve corresponding to the parameters used in \cref{fig:slices}, while the orange, green, and red curves show how varying the model parameters affects our results. As discussed below \cref{eq:ychi}, the curves are independent of the phase transition temperature $T_n$, as long as $y_\chi$ is scaled with temperature according to \cref{eq:ychi} and as long as $\SI{0.1}{TeV} \lesssim T_n \lesssim \SI{e10}{TeV}$. At higher temperatures, it is no longer possible to plot our results in a $T_n$-independent way.  Therefore, we show in the right-hand panel of \cref{fig:rho} two illustrative curves at a fixed temperature, $T_n = \SI{e12}{TeV}$. At lower temperatures, there is a dependence on $g_\star$.  To gain more insight into the dynamics of the phase transition, in \cref{fig:crosscheck} we also compare the number of $\chi$ particles inside the shrinking bubble (solid) to the number that have left the bubble and entered the true vacuum region (dashed), or have annihilated (dotted and dot-dashed). We also show the sum of these four curves (thin black).  The horizontality of this curve serves as a consistency check of our simulation, as discussed in \cref{sec:continuity}.  In \cref{fig:crosscheck} we use the same parameter points as in \cref{fig:rho}.

Focusing first on our main benchmark parameter point (solid blue curve in the left-hand panel of \cref{fig:rho} and top-left panel in \cref{fig:crosscheck}) we see that $\ev{\rho_\chi}$ increases rapidly until the bubble radius has shrunk by about a factor of three, at which point the Schwarzschild criteria is satisfied and a black hole forms (as roughly expected from \cref{eq:schwarzschild-crit}). As anticipated, the Jeans instability criteria is only satisfied shortly before the Schwarzschild criteria.  However, at this point the gravitational force is not sufficient to overcome degeneracy pressure, so increases in the energy density still rely on the compression from the wall and a black hole is not formed until the Schwarzschild criteria is satisfied.  As is evident from the top-left panel of \cref{fig:crosscheck}, black hole formation is successful because essentially all of the initial $\chi$ particles remain trapped in the bubble, with only a per cent-level fraction being lost via annihilation or escape through the bubble wall.

The solid orange curve in \cref{fig:rho} (left) shows the evolution of the average energy density for a larger bubble starting with $r_w^0 = 2\,r_H^0$. We see that the average energy density again increases rapidly until a black hole is formed. Once again, satisfying the Jeans instability criteria is not sufficient for black hole formation because gravity cannot overcome degeneracy pressure. Consequently, a black hole forms only after the Schwarzschild criteria is satisfied. The bubble radius must shrink by a factor of two before this is the case (again, as roughly expected from \cref{eq:schwarzschild-crit}). At that time the bubble radius has shrunk to around $r_w(t) \approx r_H^0$. Note, however, that in the meantime the Hubble radius has increased, so the bubble is well within the causal horizon at the time the Schwarzschild criteria is satisfied, and causality does not prevent black hole formation. More precisely, the time it takes the bubble to shrink from $r_w^0 = 2\,r_H^0$ to $r_w(t) \approx r_H^0$ is $t - t_0 \approx r_H^0 / v_w$. During this time, $r_H(t)$ grows linearly with time, implying that $r_H(t) \approx r_H^0 (1 + 1/v_w) = 3\,r_H^0$. This is larger than $2\,r_w(t) \approx 2\,r_H^0$ at the time the Schwarzschild criteria is first satisfied, so the bubble's full diameter will be inside the Hubble horizon and a black hole will quickly form.\footnote{Similar arguments can be made for all parameter points shown in this paper. For smaller $v_w$ or smaller $r_w^0$, the bubble will be even deeper inside the horizon at the time the Schwarzschild criteria is first satisfied.} In the top-right panel of \cref{fig:crosscheck} we see that for the orange parameter point nearly all $\chi$ particles once again remain trapped inside the bubble. Annihilation and losses through the bubble wall are again negligible.  The number of particles inside the bubble, $N_\chi$, increases slightly shortly before black hole formation due to the onset of gravitational collapse where our simulation becomes numerically unstable.  This does not have any impact on our conclusions.  

Bubbles whose initial radius is \emph{smaller} than $\sim 1.5\,r_H$ would need to reach an even larger overdensity before satisfying the black hole formation criteria. The immense pressure of such a large overdensity would need to be compensated by a very large latent heat release in the phase transition. The latter would violate one of our simplifying assumptions, for reasons that will be discussed in \cref{sec:latent-heat}.

The dashed green curve in \cref{fig:rho} (left) illustrates the evolution of the energy density for a smaller $\chi$ mass gain, $m_\chi^\infty = 5\,T_n$, than for the blue reference curve.  We see that the green curve flattens out before the Schwarzschild or Jeans instability criteria are satisfied. We conclude that black holes cannot form for this parameter point. The middle-left panel in \cref{fig:crosscheck} reveals that this is because, for $m_\chi^\infty = 5\,T_n$, a significant fraction of $\chi$ particles can build up sufficient momentum to escape through the wall.

Finally, the dot-dashed red line in \cref{fig:rho} (left) has been calculated under the assumption of a larger Yukawa coupling $y_\chi$.  Once again, the curve flattens before either criteria is satisfied, and the energy density does not become large enough for black hole formation.  As we can see in the middle-right panel of \cref{fig:crosscheck}, this is predominantly caused by $\chi\bar\chi \to \phi$ annihilation, which is now more efficient than for the blue, orange, and green curves.

Moving to the right-hand panel of \cref{fig:rho}, the solid brown curve shows that black hole formation can occur for high phase-transition temperatures just as well as for low ones. In fact, the brown curve is nearly identical to the blue reference curve in the left panel.

The dot-dashed purple curve, however, introduces a novel feature: the annihilation channel $\bar\chi \chi \to \phi\phi$, which is only relevant for $\mathcal{O}(1)$ Yukawa couplings $y_\chi$. In low-temperature phase transitions ($T_n \lesssim \SI{e10}{TeV}$), such large Yukawa couplings would completely prevent the build-up of a $\chi$ overdensity. At higher temperatures, the Hubble radius is much smaller, the phase transition proceeds much faster, and therefore even large Yukawa couplings can lead to black hole formation. The importance of $2 \to 2$ annihilation at the purple parameter point is seen in the bottom-right panel of \cref{fig:crosscheck}, where this annihilation process accounts for the majority of the particle loss.

We now analyse the shape of the curves in \cref{fig:rho} in more detail. For all curves, the density increase initially roughly follows the power law $\ev{\rho_\chi} \propto [r_w(t) / r_w^0]^4$. This scaling can be understood from the scaling of the number density, $\propto [r_w(t) / r_w^0]^3$, combined with the energy gain that particles experience on reflection from the moving wall, see the discussion above \cref{eq:rho-chi}. Actually, at early times, just after the start of the simulation, the density grows even faster because of relativistic corrections that are no longer negligible at the chosen wall velocity $v_w = 0.5$. In particular, the energy gain in each reflection is larger for $\gamma_w > 1$ than in the non-relativistic case. The behaviour of relatively tangential modes ($p_\sigma > p_r$) is also important. These $\chi$ particles experience two counteracting effects: on the one hand, they gain radial momentum during each reflection, so in each reflection they become less tangential and the time between reflections increases. On the other hand, they become more tangential between reflections since the wall is moving inwards. The angle of reflection is important because the energy gain in each collision is larger for head-on reflections than for grazing reflections.  All of these effects are more pronounced for faster walls. Their interplay leads to the oscillating behaviour visible in the curves in \cref{fig:rho}: at the start of the simulation, a population of once-reflected particles builds up inside the bubble, but because initially tangential modes become more radial after reflection, the number of reflections per unit time and thus the bubble's energy gain per unit time intermittently \emph{decrease}, leading to a slight flattening of the $\ev{\rho_\chi}$ curves. Once the wave of once-reflected particles has had enough time to traverse the bubble and begins to experience its second reflection, the pace of increase picks up again.

In \cref{fig:rho} we have purposefully chosen parameter points that are near to the critical values for black hole formation, to provide some insight into the physical processes at play. Smaller Yukawa couplings, larger $m_\chi^\infty/T_n$, or larger $r_w^0$ than used in the blue and orange curves also lead to black hole formation. We have thus demonstrated that bubbles initially larger than $\sim 1.5\,r_H$ with Yukawa couplings smaller than $\sim 10^{-5} \sqrt{T_n / \text{PeV}}$ and mass jumps larger than $\sim 10\,T_n$ will generate energy overdensities large enough to trigger collapse into a black hole.  In ref.~\cite{Baker:2021nyl} we discuss the resulting black hole population from the full phase transition.

Although we have studied the case of a particular toy model with a single Dirac fermion, the mechanism is expected to work in more general cases.  The new particle responsible for the overdensity could equally be a vector or scalar particle (in which case degeneracy pressure could not halt collapse after the Jeans instability criteria is satisfied).  In our model we assumed that $\phi$ was in equilibrium with the SM bath throughout the phase transition.  If this is not the case then the phase space distribution function of $\phi$ would also need to be tracked.  While  we assumed that the Universe was radiation dominated when relating the Hubble parameter to the temperature in our estimates in \cref{sec:Schwarzschild-criteria,sec:JI-criteria}, the mechanism is also expected to work in phases of matter domination.  Finally, although we have studied the case where $\chi$ obtains a large mass due to a Higgs mechanism, the mechanism could also work in other situations, for instance during a confining phase transition.

\section{Discussion}
\label{sec:discussion}

We now discuss in detail the conditions under which our mechanism is expected to work and the assumptions that go into our calculations. We will address in particular constraints on the strength of the phase transition (expressed through the latent heat release $\Delta V$), as well the evolution of the wall velocity and wall thickness.  We will see that physically we do not expect the wall velocity to remain constant.  Instead, we would expect the wall to initially accelerate, before decelerating again due to the increasing pressure inside.  However, we will see that our use of a constant wall velocity in simulations is typically conservative (in the sense that in many models black hole formation would be successful in a wider range of scenarios than those we demonstrate).

\subsection{Required Strength of the Phase Transition}
\label{sec:latent-heat}

An obvious precondition for successful black hole formation is that the bubble can keep shrinking while a sufficient $\chi$ overdensity builds up.  This implies that the pressure that drives the bubble wall forward, $P_V = \Delta V$, should be greater than the pressure from particles interacting with the wall. Here, $\Delta V$ denotes the effective potential difference between the true and false vacua (that is, the latent heat release associated with the phase transition). The pressure from particle interactions will be dominated by interactions of $\chi$ particles, $P_\chi$, since all other particle species are equally abundant on both sides of the wall.  The pressure $P_\chi$ is given by~\cite{Megevand:2009gh,Moore:1995si}
\begin{align}
    P_\chi(t) &= g_\chi \int\!\upd r \int\!\frac{\upd^3p}{(2\pi)^3}
                 \pd{E}{m_\chi} \pd{m_\chi}{r} \delta \f \,,
    \label{eq:pressure}
\end{align}
where $\delta f_\chi \equiv f_\chi - f_\chi^\text{eq}$ is the deviation of the phase space distribution function $f_\chi(r, p_r, p_\sigma, t)$ from an equilibrium distribution $f_\chi^\text{eq}(r, p_r, p_\sigma, t)$, and $E(r, p_r, p_\sigma, t)$ is the energy of a $\chi$ particle. Note it is only the $\chi$ overdensity that enters $P_\chi$, not the total density, because the contribution from an equilibrium population of $\chi$ is already accounted for by the effective thermal potential difference $\Delta V$~\cite{Moore:1995ua}. An equivalent way of stating the condition
\begin{align}
    P_\chi(t) < P_V = \Delta V \,
    \label{eq:pressure-cond}
\end{align}
is to say that the work which a surface element $\upd A$ of the wall has to perform against the pressure $P_\chi$ to advance a distance $\upd r$ should be smaller than the potential energy released due to the corresponding expansion of the true-vacuum phase.

However, $\Delta V$ should not be arbitrarily large. If the latent heat of the phase transition is greater than the energy density due to radiation, then the interior of the bubble, where the false vacuum persists, will be vacuum energy dominated and the scale factor will undergo a period of exponential growth.  The $\chi$ particles inside the bubble will then be significantly red-shifted, reducing the energy density.  This will make it harder for black holes to form.  Although this does not necessarily mean that black holes cannot form, for simplicity we require that the latent heat is small enough so that there is no period of vacuum energy domination,
\begin{align}
    \Delta V \lesssim &\, \rho_\text{rad}(t_0) = \frac{\pi^2}{30} g_\star T_n^4 \,,
        \label{eq:DV-condition}
\end{align}
where $g_\star$ is again the effective number of relativistic degrees of freedom, including $\chi$ and $\phi$.  Since $\rho_\chi$ grows as the bubble shrinks, satisfying the condition $\Delta V \lesssim \rho_\text{rad}(t_0)$ ensures that there is radiation domination at all times.

Taking the two conditions together, we require that the pressure due to $\chi$ is less than the initial radiation energy density,
\begin{align}
    P_\chi(t) < \rho_\text{rad}(t_0)\,,
    \label{eq:Pchi-condition}
\end{align}
at all times, and assume that the phase transition has a latent heat satisfying $P_\chi(t) < \Delta V < \rho_\text{rad}(t_0)$. 

Quantitatively, \cref{eq:Pchi-condition} is indeed satisfied for the successful parameter points presented in \cref{fig:rho}.  For the blue and brown lines, with $r_w^0 = 1.5\,r_H^0$, we find $P_\chi/\rho_\text{rad}(t_0) \lesssim 0.6$ at all times, so $\Delta V$ does not need to be fine tuned.  For the orange curve, with $r_w^0 = 2\,r_H^0$, an even wider range of $\Delta V$ is allowed since $P_\chi/\rho_\text{rad}(t_0) \lesssim 0.03$.  This ratio is so small since the main wave of once-reflected particles has not had time to traverse the bubble by the time a black hole forms, so the pressure due to $\chi$ remains low throughout. As mentioned in \cref{sec:results}, initial bubble radii $r_w^0 \gtrsim 1.5\,r_H^0$ are required to avoid large energy overdensities which would in turn imply a period of inflation, violating our simplifying assumption.\footnote{Note that assuming a larger radiation density (due to extra degrees of freedom beyond the SM) would not solve this problem as it would also make the Schwarzschild criteria more stringent, see \cref{eq:schwarzschild}.}

\subsection{Wall Velocity and Wall Thickness}
\label{sec:dependence-wall-velocity}

In our numerical simulation we assumed the bubble wall to advance at a constant velocity $v_w$. However, the discussion above, in particular \cref{eq:pressure-cond}, suggests that physically this will not generally be the case: the pressure pushing the wall forward, $P_V = \Delta V$, remains constant throughout the phase transition, while the pressure in front of it, $P_\chi$, increases as a $\chi$ overdensity gradually builds up.  This leads to potentially significant (and highly model-dependent) changes in $v_w$ during the build-up of the $\chi$ overdensity. Moreover, Hubble expansion also modifies $v_w$ by counteracting the advance of the bubble wall. Fortunately, as we will discuss below, our conclusions are conservative for relativistic walls (as long as $\gamma_w \ll m_\chi^\infty/T_n$) and nearly independent of $v_w$ for moderate wall velocities $0.1 \lesssim v_w \lesssim 0.5$. The case of very small $v_w \ll 0.1$ is of little practical interest as it would require substantial tuning between the latent heat release that drives the wall forward and the retarding forces, in particular $P_\chi$.  We will also argue that our results are independent of the wall thickness $l_w$, as long as $l_w \ll r_w$.

After the onset of the phase transition, the latent heat release will typically accelerate the bubble wall.  Based on dimensional grounds, we estimate that the wall accelerates over distance scales of order $T_n^3 / \Delta V$. Assuming $T_n \ll M_\text{Pl}$ and $\Delta V \lesssim (\pi^2/30) g_\star T_n^4$, to avoid a period of vacuum domination, this distance scale is much shorter than the distance $\sim 1/H \sim M_\text{Pl}/T_n^2$ which we have found the wall must cover to accumulate an overdensity large enough to trigger black hole formation.  This means that the bubble wall typically accelerates to relativistic speeds at the very start of the compression.

For relativistic wall speeds, the momentum gain that $\chi$ particles experience upon being reflected off the wall is significantly larger than for non-relativistic walls, which implies that the energy overdensity inside the bubble grows even faster than for non-relativistic wall speeds.  This will make black hole formation even more likely. Since our implementation is only numerically stable for $v_w \lesssim 0.5$ (we take $v_w = 0.5$ in our benchmark scenarios), this means that we will overestimate the amount of compression required to form a black hole in scenarios where $v_w \sim 1$.  Since our aim is to show that black hole formation is possible via this mechanism, our results for $v_w = 0.5$ are a conservative estimate for scenarios where $v_w \sim 1$ (as long as $\gamma_w \ll m_\chi^\infty / T_n$).

If the wall were able to reach a boost $\gamma_w \gtrsim m_\chi^\infty / T_n$ (which we here call `ultra-relativistic'), it would become transparent to $\chi$ particles because then the average $\chi$ energy, which in the wall's rest frame is of order $\gamma_w T_n$, would be large enough to allow $\chi$ particles to enter in spite of their large mass gain. In this case, no overdensity and thus no black holes could form.  One possibility of avoiding this is if friction due to the reflection of $\chi$ particle is large enough to prevent acceleration of the wall to ultra-relativistic velocities already at the beginning of the compression, when $\ev{\rho_\chi} \simeq \rho_\chi^\text{eq}$. However, as friction increases further as the bubble shrinks and a $\chi$ overdensity builds up, there is a risk that the wall will slow and and stop before the overdensity becomes large enough to allow a black hole to form. This fate is avoided if most of the $\chi$ particles are only reflected by the wall once before black hole formation because the wall will then not feel the pressure due to the $\chi$ overdensity.  Most of the $\chi$ particles will only be reflected once if relatively little compression is required, as in the orange benchmark point in \cref{fig:rho}, or if the wall is moving with $v_w \sim 1$.  For bubble walls moving nearly at the speed of light, there is simply not enough time for multiple reflections to occur before the bubble shrinks to zero

Another possibility for avoiding ultra-relativistic wall velocities is if additional friction forces are at play. This is, in fact, typically the case in complete models, where other new fields (in addition to $\phi$ and $\chi$) are present. In particular, transition radiation emitted by particles traversing the bubble walls can be crucial~\cite{Bodeker:2017cim, Hoche:2020ysm, Azatov:2020ufh, Long:2024sqg}.  We imagine for instance the presence of an extra massive vector boson $A'$ that can be radiated off particles that pass through the bubble wall.  The retarding pressure due to this transition is $P_{A'} \sim \gamma_w g^2 m_{A'} T_n^3$, where $m_{A'}$ is the mass of $A'$ and where $\gamma_w$ is the Lorentz boost of the wall~\cite{Bodeker:2017cim,Long:2024sqg}.  This can be efficient if $m_{A'} \lesssim m_\chi^\infty$ and $g \sim 1$.  Assuming $m_\chi^\infty \sim 10\,T_n$, then before the bubble wall reaches $\gamma_w \sim m_\chi^\infty / T_n$ -- the critical value beyond which the $\chi$ particles can escape through the bubble wall -- transition radiation becomes dominant and generates a pressure that is sufficient to counteract $P_V$. This will prevent the wall from accelerating further, so the $\chi$ overdensity continues to grow, slowing down the wall again. Eventually, transition radiation will become subdominant again and it will be the $\chi$ overdensity alone that balances $P_V$.  We see in this example the strong model-dependence of the wall velocity.

\begin{figure}
    \centering
    \includegraphics{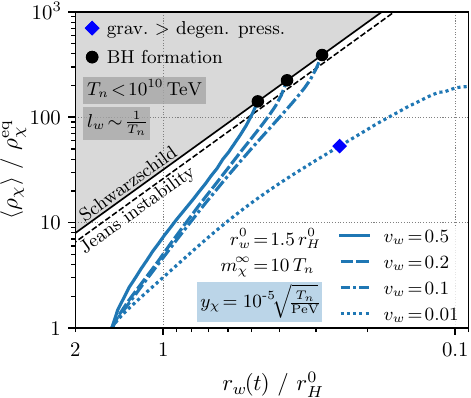}\hfill
    \includegraphics{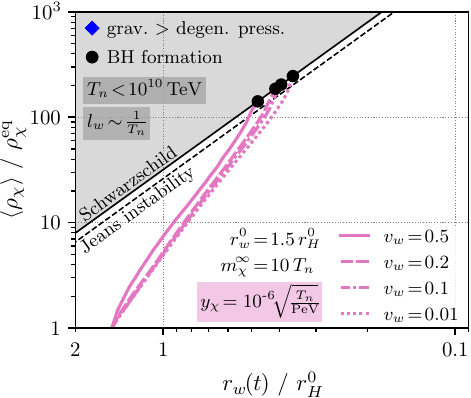}
    \caption{
    The average energy density of $\chi$, $\ev{\rho_\chi}$, as a function of radius for different wall velocities $v_w$.
    }
    \label{fig:rho-wall-velocity}
\end{figure}

It remains to demonstrate that our numerical results are independent of the wall velocity $v_w$ for $0.1 \lesssim v_w \lesssim 0.5$. In \cref{fig:rho-wall-velocity} we show the $\chi$ energy density averaged over the bubble as a function of the bubble radius for several wall velocities.  The curves in the left-hand panel correspond to the same parameter point as the blue curve in \cref{fig:rho}, but with $v_w \in \{0.5, 0.2, 0.1, 0.01 \}$.  Comparing the curves, we see marginal differences from changing $v_w$ from $0.5$ to $0.2$ or $0.1$. Larger $v_w$ leads to somewhat earlier black hole formation because the energy gain in each reflection is larger.  When the wall velocity is $\ll 0.1$, there is more time for particles to annihilate before a black hole can form, which is why the dotted blue curve in the left-hand panel of \cref{fig:rho-wall-velocity} never reaches the threshold for black hole formation. This problem can be avoided by choosing a smaller $y_\chi$, as in the right-hand panel of \cref{fig:rho-wall-velocity}.

We have so far neglected Hubble expansion, which stretches the space inside the bubble while the advancing bubble wall strives to shrink the bubble.  It does not, however, change the relative velocity at which the wall travels through the plasma, so the dynamics in the vicinity of the bubble wall are unaffected. Only the time it takes the bubble to shrink to a given radius becomes larger, but since we show our results as a function of $r_w(t) / r_H^0$, our plots would remain unchanged.  Hubble expansion also leads to redshift, which we neglect as we have focused on scenarios in which black holes form in not much more than one Hubble time. (At small wall velocities, $v_w \ll 0.5$, this approximation may no longer be justified, and Hubble expansion during the phase transition would need to be included in our calculations.) 

Finally, we note that our results are independent of the wall width $l_w$, as long as $l_w \ll r_w$. This holds when $T_n \ll M_\text{Pl}$ since we have $l_w \sim 1/T_n$ and $r_w \sim r_H \sim M_\text{Pl}/T_n^2$. As discussed in \cref{sec:liouville}, the Boltzmann equation can then be approximated in different ways in the bulk of the bubble and near the wall.  While $l_w$ affects the phase space trajectories close to the wall, it does not change the way $\chi$ particles are reflected or transmitted, when viewed far from the wall.  Since black hole formation depends only on quantities in the bulk of the bubble, it is independent of the wall width.

\subsection{Black Hole Formation Probability}
\label{sec:formation-probability}

We now consider how likely it is that no true-vacuum bubble nucleates inside a large false-vacuum bubble, which can tell us about the expected abundance of primordial black holes. Answering this question in full detail is beyond the scope of this work as it would require hydrodynamic simulations of the phase transition to determine the distribution of initial radii, $r_w^0$, of the false-vacuum pocket as well as the probability that the pocket survives long enough to form a black hole before a new true vacuum bubble nucleates inside. We can, however, make some estimates to demonstrate that a sizable black hole formation rate can be achieved.

\begin{figure}
    \centering
    \includegraphics[width=0.5\textwidth]{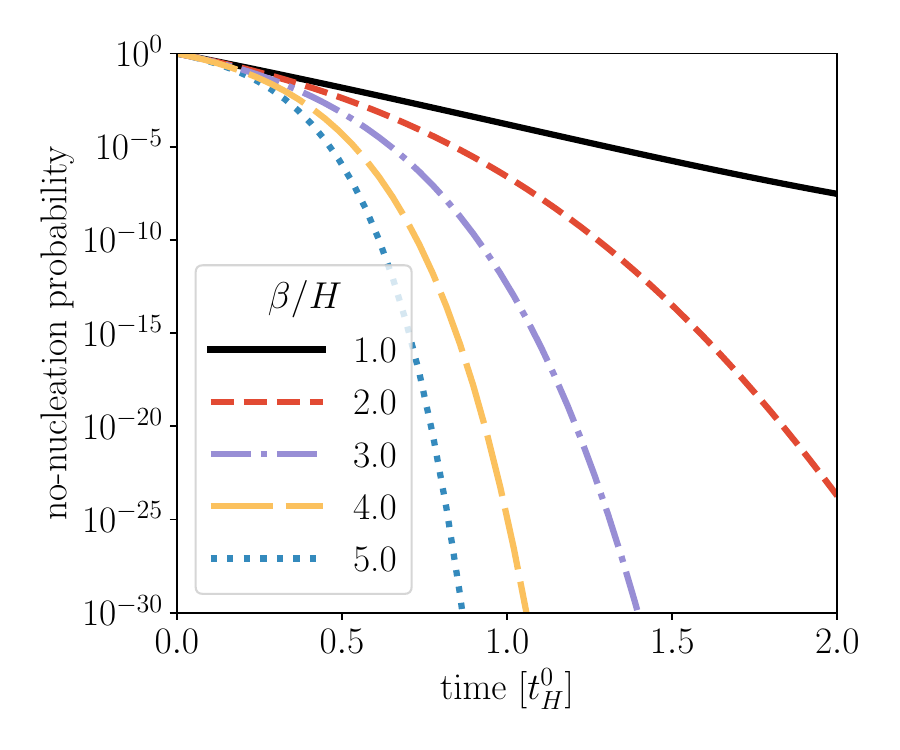}
    \caption{The probability as a function of time of \emph{not} nucleating a true-vacuum bubble in a shrinking false-vacuum pocket. Different curves corresponds to phase transitions of different speed $\beta$.}
    \label{fig:p-no-nuc}
\end{figure}

We first consider a spherical false-vacuum pocket of initial radius $r(t_0) = r_0$ that is shrinking isotropically with wall velocity $v_w$, such that its radius at time $t$ is $r(t) = r_0 - v_w (t - t_0)$. The bubble nucleation rate is (see e.g.\ \cite{Megevand:2017vtb})
\begin{align}
    \Gamma_\text{nuc}(t) = H \big[r(t) H \big]^3 e^{\beta (t - t_0)} \,,
    \label{eq:Gamma-nuc}
\end{align}
where we have assumed that the nucleation rate at $t=t_0$ is one bubble per Hubble time per Hubble volume. We treat the speed of the phase transition, $\beta$, as an free parameter here, but in a full model it would be related to the temperature, $T$, and the Euclidean bounce action, $S_E(T)$, via $\beta / H = T \, S'_E(T)$. The probability that \emph{no} true-vacuum bubble has nucleated inside the false-vacuum pocket after a time $t$ is
\begin{align}
    p_\text{no-nuc} = \exp\Big[ -\int_0^t \! \upd t' \Gamma_\text{nuc}(t') \Big]
\end{align}
This quantity is shown in the left panel of \cref{fig:p-no-nuc} for several choices of $\beta$. For simplicity, we neglect the time dependence of the Hubble rate. We see that $p_\text{no-nuc}$ drops exponentially as the system evolves. However, for slow phase transitions (small $\beta$), the probability of not nucleating a bubble inside the shrinking false-vacuum region (but instead forming a primordial black hole) is non-negligible. Note that even accounting for all of the dark matter in the Universe requires only $p_\text{no-nuc} \sim \mathfrak{p} = \num{e-20}$--\num{e-15}~\cite{Baker:2021nyl}.  In the notation of ref.~\cite{Baker:2021nyl}, $\mathfrak{p}$ is the probability of forming a primordial black hole in a given Hubble volume. It is similar to $p_\text{no-nuc}$ except in situations where $\chi$ annihilation, or loss of $\chi$ particles through the bubble wall, inhibits black hole formation.

\section{Conclusions}
\label{sec:conclusions}

We have presented a detailed discussion of black hole formation during cosmological first-order phase transitions due to the build-up of particles which obtain a large mass during the phase transition.  We have described the bulk properties of the phase transition required for the general mechanism to work, and the simplifying assumptions we make in our numerical simulations.  We have discussed in detail the criteria for forming a black hole, of which the most important is the Schwarzschild criteria.  We have discussed the Boltzmann equation for the phase space distribution function of the affected particles and outlined the steps we take and approximations we make to perform our numerical simulations.  Finally we have presented and discussed the results of our simulation, highlighting the regions of parameter space that lead to successful black hole formation.

The mechanism requires relatively few ingredients to potentially form black holes.  There needs to be a first-order phase transition, which could be driven by either a scalar or a confining sector.  There also needs to be a particle whose mass in the false vacuum state is significantly smaller than the one in the true vacuum.  As the phase transition completes, these particles are confined to small regions, leading to large energy overdensities that may collapse to form black holes.  To demonstrate that black holes may form, we make further assumptions that, while not strictly required for the mechanism to work, simplify our numerical simulations.  We assume that the phase transition is independent of the electroweak phase transition, but that the new scalar remains in thermal contact with the Standard Model bath.  We further assume that the shrinking regions of false vacuum are spherically symmetric and that the wall profile, velocity and thickness are constant throughout the phase transition.  Of these, the constant wall velocity is perhaps the hardest to justify.  However, we have demonstrated that black hole formation is mostly independent of the actual wall velocity, as long as the wall's relativistic gamma factor is less than the mass gain divided by the temperature.  We finally assume that the Universe remains radiation-dominated throughout.

We showed that the key criteria for black hole formation is the Schwarzschild criteria (that is, the requirement that the region containing the energy overdensity becomes smaller than its Schwarzschild radius).  Using this criteria, we saw that simple estimates could provide a guide as to what to expect from the full numerical simulation.  We also considered weaker black hole formation criteria, where the overdensity forms a Jeans instability and overcomes degeneracy pressure.  However, we found that these conditions do not typically lead to earlier black hole formation than the Schwarzschild criteria.

To make the Boltzmann equations tractable, we exploited the approximate spherical symmetry to remove two spatial and one momentum component from the problem, simplifying the Liouville operator.  In the collision term we included the dominant effects due to annihilation.  We then numerically solved the Boltzmann equation using the method of characteristics.  This method also provides an intuitive picture of the trajectories of particles, in the absence of collisions.

Finally we presented our numerical results.  We described in detail the overdensity at different times for different slices in momentum space for a particular choice of parameters, showing the energy overdensity building up inside the shrinking bubble.   We then varied the parameters, showing in \cref{fig:rho} that, assuming a large spherical vacuum pocket of sufficient size and a constant wall velocity, black holes could be formed for large initial bubbles, $r_w^0 \gtrsim 1.5\,r_H^0$, and for large mass gains, $m_\chi^\infty \gtrsim 10\,T_n$, for a wide range of Yukawa couplings and wall velocities.

\section*{Acknowledgements}

It is a pleasure to thank Xiaoyong Chu, Valerie Domcke, Ivan Esteban, Juri Smirnov, Azadeh Maleknejad, Alexander Zhiboedov for interesting and insightful discussions.
MJB would like to acknowledge support from the Australian Government through the Australian Research Council.
The work of MB, JK, and LM has been partly funded by the German Research Foundation (DFG) in the framework of the PRISMA+ Cluster of Excellence and by the European Research Council (ERC) under the European Union's Horizon 2020 research and innovation programme (grant agreement No.\ 637506, ``$\nu$Directions''). MB and JK would also like to acknowledge support from DFG Grant No.~KO-4820/4-1.

\bibliographystyle{JHEP}
\bibliography{refs}

\end{document}